\documentstyle[12pt]{article}
\newcommand{\bra}{\langle}
\newcommand{\ket}{\rangle}
\newcommand{\di}{\mbox{d}}
\newcommand{\e}{\mbox{e}}
\newcommand{\k}{\mbox{i}}
\newcommand{\sk}{\mbox{\scriptsize i}}
\newcommand{\bm}[1]{\mbox{\boldmath{$ #1 $}}}
\newcommand{\B}{{\cal B}}
\newcommand{\F}{{\cal F}}
\newcommand{\g}{{\cal G}}
\newcommand{\h}{{\cal H}}
\newcommand{\hp}{{\cal H}^\perp}
\newcommand{\so}{\mbox{\boldmath{$ so $}}}
\newcommand{\su}{\mbox{\boldmath{$ su $}}(2)}
\newcommand{\Lieu}{\mbox{\boldmath{$ u $}}(1)}
\newcommand{\Lb}{L}
\newcommand{\A}{{\cal A}}
\newcommand{\Hil}{\Gamma}
\newcommand{\Gm}{\Gamma}
\newcommand{\Gs}{\Gamma_\sigma}
\newcommand{\Gn}{\Gamma_n}
\newcommand{\Ga}{\Gamma_\alpha}
\newcommand{\Es}{E_\sigma}
\newcommand{\pis}{\pi_\sigma}
\newcommand{\cs}{c_\sigma}
\newcommand{\Om}{\Omega}
\newcommand{\CP}{\mbox{\boldmath{$C$}}P^1}
\newcommand{\RP}{\mbox{\boldmath{$R$}}P^n}

\newenvironment{enum1}%
{\begin{enumerate}}%
{\end{enumerate}}
\newenvironment{enum2}%
{\begin{enumerate}}%
{\end{enumerate}}
\makeatletter
\@addtoreset{equation}{section}
\makeatother

\begin{document}
%
%
\begin{flushright}
	{\tt DPNU-93-21
	\\
	June 1993}
\end{flushright}
\vspace*{24mm}
\begin{center}
	{\LARGE\bf
	Quantum Mechanics on Manifolds
	}
	\\
	\vspace{12mm}
	{\large Shogo {\sc Tanimura}}
	\footnote{e-mail address : tanimura@eken.phys.nagoya-u.ac.jp}
	\\
	\vspace{6mm}
	{\large \it Department of Physics, Nagoya University, }
	\\
	\vspace{4mm}
	{\large \it Nagoya 464-01, Japan}
\\
\vspace{24mm}
{\bf Abstract}
\\
\vspace{6mm}
\begin{minipage}[t]{130mm}
\baselineskip 6mm
	A definition of quantum mechanics on a manifold $ M $ is proposed
	and a method to realize the definition is presented.
	This scheme is applicable to a homogeneous space $ M = G / H  $.
	The realization is
	a unitary representation of the transformation group $ G $
	on the space of vector bundle-valued functions.
	When $ H \ne \{ e \} $,
	there exist a number of inequivalent realizations.
	As examples, quantum mechanics on
	a sphere $ S^n $, a torus $ T^n $ and a projective space $ \RP $
	are studied.
	In any case, it is shown that
	there are an infinite number of inequivalent realizations.
\end{minipage}
\end{center}
\newpage
\baselineskip 7mm
%
%
\section{Introduction}
The geometrical approach to quantum mechanics has been developed
to be an important branch of mathematical physics.
Geometry is a study of properties of a space
which are invariant under action of a transformation group.
For instance,
coordinates of a point on a manifold vary under transformation of coordinates,
hence coordinates themselves are not direct objects of geometry.
They are rather artificial objects.
On the other hand, as everyone knows,
physics is a study of properties of nature
which are invariant under changes of observers.
Therefore laws of physics should be expressed in terms of geometry.
Even quantum mechanics cannot be an exception.
\par
Many authors have been investigating
the geometrical approach to quantum mechanics.
Bayen et al.~\cite{Bayen} and Batalin and Tyutin~\cite{Batalin} have taken
a phase space $ M $ as a base space.
Both of them have treated
well-defined complex-valued functions on the phase space $ M $.
A common feature of their formulations is
introducing associative but noncommutative multiplication
among functions on $ M $, which is called $ \ast$-multiplication.
Both have constructed an algebra of functions by $ \ast$-multiplication
which is isomorphic to the algebra of quantum-mechanical operators.
Furthermore, Bayen et al.~\cite{Bayen} have calculated
spectra of physical quantities
such as energies of the harmonic oscillator and the hydrogen atom
{\it without introducing a Hilbert space}.
That is in fact a surprising result.
They have obtained discrete spectra
using only classical-mechanical functions on the phase space
and $ \ast$-multiplication.
\par
However, we can also take
a configuration space instead of a phase space as a base.
Recently, Ohnuki and Kitakado~\cite{OK1}, \cite{OK2} have considered
the case where the configuration space is a sphere $ S^n $,
and formulated quantum mechanics on $ S^n $.
They~\cite{OK2} have defined
the fundamental algebra $ \A $ of quantum mechanics on $ S^n $,
which is a substitution for
the canonical commutation relations of quantum mechanics on $ \bm{R}^n $.
Moreover, they have defined quantum mechanics on $ S^n $
as an irreducible representation of $ \A $.
They have shown that
{\it there exist an infinite number of inequivalent representations}.
That is a noticeable result.
Existence of inequivalent representations means existence of different physics.
Each representation gives different evaluation to a physical quantity,
for example,
spin~\cite{OK1}, \cite{OK2}, probability amplitude and energy~\cite{Tani}.
\par
In this paper we take a configuration space $ M $ as a base space.
The purpose of this paper is
to propose a definition of quantum mechanics on $ M $
and to present a method to realize the definition.
Considered manifolds are homogeneous spaces, which are defined in the text.
In section~2, we define quantum mechanics on $ M $
and present a method to construct it.
In section~3, we study some examples, that is, quantum mechanics on
a sphere $ S^n $, a torus $ T^n $ and a projective space $ \RP $.
In any case,
we notice that there are an infinite number of inequivalent realizations.
Section~4 is devoted to discussions.
There we give physical interpretation to our formulation.
\par
The readers are assumed to be familiar with differential geometry
at the level of the literature~\cite{Eguchi}, \cite{Nakahara}.
We use mathematical terminology obeying the dictionary~\cite{Dictionary}.
%
%
\section{Definition and Construction}\label{sect:Def}
\subsection{Definition of quantum mechanics on a manifold}%
\label{sect:definition}
First, we propose a definition of quantum mechanics on a manifold.
We prefer to consider as general manifolds as possible.
In this paper we consider a manifold $ M $
possessing the following structure $ ( G, \tau, \B, \mu ) $:
\begin{enum1}
\item	$ G $ is a transformation group acting on $ M $ transitively.
	We shall give details of this statement.
	$ G $ is a Lie group.
	$ \tau $ is a differentiable map
	\begin{equation}
		\tau : G \times M \to M , \qquad ( a, x ) \mapsto a x
		\label{eqn:2.1}
	\end{equation}
	which satisfies the conditions:
	\begin{eqnarray}
		&&
		( a b ) x = a ( b x ), \qquad a, b \in G, \quad x \in M,
		\label{eqn:2.2}
		\\
		&&
		e x = x, \qquad \qquad \quad
		e \in G (\mbox{unit element}), \quad x \in M.
		\label{eqn:2.3}
	\end{eqnarray}
	$ \tau $ is called an action of $ G $ on $ M $.
	The action is said to be transitive if,
	for arbitrary two points $ x, y \in M $,
	there exists an element $ a \in G $ such that $ a x = y $.
	The manifold $ M $ admitting the transitive action of $ G $ is called
	a homogeneous space.
\item	$ \B $ is a topological $ \sigma $-algebra of $ M $.
	$ \mu $ is a $ G $-invariant measure on $ \B $.
	The set $ ( M, \B, \mu ) $ forms a Borel measure space.
	$ \F(M) $ denotes
	a space which consists of
	$ \B $-measurable complex-valued functions.
	A detailed explanation of the terminology
	is found in the dictionary~\cite{Dictionary}.
\end{enum1}
We define quantum mechanics on $ M $ as a set $ ( \Gm, \nu, \rho, H ) $
which consists of the following:
\begin{enum1}
\item	$ \Gm $ is a Hilbert space.
\item	$ \nu $ is a map
	\begin{equation}
		\nu : \Gm \times \Gm \to \F(M),
		\qquad
		( \varphi, \psi ) \mapsto \nu( \varphi, \psi ),
		\label{eqn:2.4}
	\end{equation}
	where $ \nu( \varphi, \psi ) $ is a complex-valued function on $ M $.
	Furthermore, $ \nu $ satisfies the conditions:
	\begin{enum2}
	\item	integral representation of inner product:
		\begin{equation}
			\bra \varphi, \psi \ket
			=
			\int_M \nu( \varphi, \psi )(x) \, \di \mu (x),
			\label{eqn:2.5}
		\end{equation}
		where the left-hand side is
		an inner product in the sense of the Hilbert space
		and the right-hand side is
		an integration with respect to the measure $ \mu $.
	\item	linearity:
		\begin{eqnarray}
			\nu( \varphi, \lambda_1 \psi_1 + \lambda_2 \psi_2)
			=
			\lambda_1 \, \nu( \varphi, \psi_1 ) +
			\lambda_2 \, \nu( \varphi, \psi_2 ),
			&&
			\nonumber
			\\
			\lambda_1, \lambda_2 \in \bm{C},
			\qquad
			\varphi, \psi_1, \psi_2 \in \Gm.
			&&
			\label{eqn:2.6}
		\end{eqnarray}
	\item	hermiticity:
		\begin{equation}
			\nu( \psi, \varphi ) = \nu( \varphi, \psi )^* .
			\label{eqn:2.7}
		\end{equation}
	\item	non-negativeness:
		\begin{equation}
			\nu( \psi, \psi )(x) \ge 0, \qquad x \in M.
			\label{eqn:2.8}
		\end{equation}
	\item	localizability:
		for an arbitrary $ D \in \B $ such that $ \mu(D) \ne 0 $,
		there exists an element $ \chi_D \in \Gm $ such that
		$ \chi_D \ne 0 $ and
		\begin{equation}
			\nu( \chi_D, \chi_D ) (x) = 0,
			\qquad
			x \notin D
			\label{eqn:2.9}
		\end{equation}
	\end{enum2}
\item	$ \rho $ is a unitary representation of $ G $ on $ \Gm $.
	For an element $ a \in G $,
	$ \rho(a) $ is a unitary operator on $ \Gm $.
	Furthermore, $ \rho $ satisfies the local unitarity condition:
	\begin{equation}
		\nu( \rho(a) \varphi, \rho(a) \psi) (a x)
		=
		\nu( \varphi, \psi) (x),
		\qquad
		x \in M.
		\label{eqn:2.10}
	\end{equation}
\item	$ H $ is a self-adjoint operator on $ \Gm $,
	which is called Hamiltonian.
	We are often interested in the $ G $-invariant Hamiltonian,
	which satisfies
	\begin{equation}
		\rho(a) \, H \, \rho(a)^\dagger = H,
		\qquad
		a \in G.
		\label{eqn:2.11}
	\end{equation}
\end{enum1}
\subsection{Preparation}
In this section we shall consider structure built in the manifold $ M $
to prepare for construction of quantum mechanics.
In what follows we will see that $ M $ inherits geometric structures
such as principal fiber bundle, Riemannian metric,
Riemannian submersion, invariant measure and connection.
\subsubsection{Principal fiber bundle}
We already have the action $ \tau $ of the group $ G $ on $ M $.
For a point $ x \in M $,
\begin{equation}
	H_x := \{ a \in G \, | \, a x = x \}
	\label{eqn:2.12}
\end{equation}
is to be a subgroup of $ G $.
We call $ H_x $ the isotropy group of $ x $.
It is also called the little group or stabilizer of $ x $.
\par
As the action $ \tau $ is transitive, all isotropy groups are conjugate.
That is to say, for arbitrary two points $ x, y \in M $,
there exists an element $ a \in G $ such that $ a x = y $.
Thus $ H_y = a \, H_x \, a^{-1} $.
\par
Now we choose a point $ p \in M $ arbitrarily, and fix it in what follows.
We put
$ H := H_p $.\footnote[5]{$ H $ is also used to denote the Hamiltonian.
However, in the following, confusion will be avoided by context.}
Introduce a relation $ \sim $ for $ u, u' \in G $
by $ u \sim u' $ if $ u p = u' p $.
$ u \sim u' $,
if and only if there exists an element $ h \in H $ such that $ u h = u' $.
The relation $ \sim $ in $ G $ is an equivalence relation.
We denote the equivalence class
$ [u] = \{ u' \in G \, | \, u \sim u' \} $
by $ u H $.
The quotient space is denoted by $ G/H $.
We define a map
\begin{equation}
	\pi : G \to M, \qquad u \mapsto u p.
	\label{eqn:2.13}
\end{equation}
Apparently, $ \pi^{-1} (x) = u H $ when $ x = \pi(u) $.
Accordingly, we can identify $ G/H $ with $ M $.
\par
On the other hand,
we define an action of $ h \in H $ on $ G $ from the right by
\begin{equation}
	R_h : G \to G, \qquad u \mapsto u h,
	\label{eqn:2.14}
\end{equation}
which satisfies $ \pi \circ R_h = \pi $.
Therefore, the set $ ( G, \pi, M, H ) $ forms a principal fiber bundle,
which consists of the total space $ G $, the base space $ M $,
the projection $ \pi $ and the structure group $ H $.
\subsubsection{Riemannian metric}
Let $ a $ and $ u $ be elements of $ G $.
The left-translation $ L_a : G \to G $ and
the right-translation $ R_a : G \to G $ of $ u $ by $ a $ are defined by
$ L_a u := au $ and $ R_a u := ua $ respectively.
$ L_a $ and $ R_a $ induce differential maps $ L_{a*} $ and $ R_{a*} $.
Let $ \g $ be the Lie algebra of the Lie group $ G $.
We identify $ \g $ with a tangent vector space of $ G $ at $ e $,
which is denoted by $ T_e G $.
The Maurer-Cartan form $ \theta : T G \to \g $ is defined by
\begin{equation}
	\theta( X ) := ( L_{u*} )^{-1} ( X ),
	\qquad X \in T_u G,
	\label{eqn:2.15}
\end{equation}
where $ T_u G $ is a tangent vector space of $ G $ at $ u $.
Furthermore,
we assume that there exists an adjoint-invariant metric
$ \beta $ of $ \g $.
Adjoint-invariance refers to the condition
\begin{equation}
	\beta( a A a^{-1}, a B a^{-1} ) = \beta( A, B ),
	\qquad
	A, B \in \g,
	\quad a \in G
	\label{eqn:2.16}
\end{equation}
or
\begin{equation}
	\beta( [ C, A ], B ) + \beta( A, [ C, B ] ) = 0,
	\qquad
	A, B, C \in \g.
	\label{eqn:2.17}
\end{equation}
The combination of $ \theta $ and $ \beta $
defines a metric $ g $ on $ G $ by
\begin{equation}
	g( X, Y ) := \beta( \theta( X ), \theta( Y ) ),
	\qquad X, Y \in T_u G.
	\label{eqn:2.18}
\end{equation}
It is easily verified that the metric $ g $ is both-invariant:
\begin{eqnarray}
	&&
	L_a^* \, g = g,
	\label{eqn:2.19}
	\\
	&&
	R_a^* \, g = g,
	\label{eqn:2.20}
\end{eqnarray}
where $ L_a^* $ and $ R_a^* $ are pullbacks induced by $ L_a $ and $ R_a $
respectively.
\par
The projection $ \pi : G \to M $ induces a differential map
$ \pi_* : TG \to TM $.
The vertical subspace $ V_u $ is a subspace of $ T_u G $
defined as the kernel of $ \pi_* : T_u G \to T_{\pi(u)} M $.
The metric $ g $ determines orthogonal decomposition of $ T_u G $
into $ V_u \oplus W_u $.
$ W_u $ is the orthogonal complement of $ V_u $
and is called the horizontal subspace.
Here we repeat the definitions:
\begin{eqnarray}
	&&
	V_u
	:= \{ X \in T_u G \, | \, \pi_* ( X ) = 0 \},
	\label{eqn:2.21}
	\\
	&&
	W_u
	:= \{ X \in T_u G \, | \,
	g( X, Y ) = 0, \quad \forall \: Y \in V_u \}.
	\label{eqn:2.22}
\end{eqnarray}
Define $ l_a : M \to M \, ( a \in G ) $ by $ l_a x := ax $.
Since $ \pi \circ L_a = l_a \circ \pi $,
$ L_{a*} V_u = V_{au} $.
Furthermore, since $ \pi \circ R_h = \pi $,
$ R_{h*} V_u = V_{uh} $.
Moreover, we have already known that the metric $ g $ is both-invariant
(\ref{eqn:2.19}), (\ref{eqn:2.20}).
Accordingly, we conclude that
\begin{eqnarray}
	&&
	L_{a*} \, W_u = W_{au},
	\label{eqn:2.23}
	\\
	&&
	R_{h*} \, W_u = W_{uh}.
	\label{eqn:2.24}
\end{eqnarray}
A restriction of $ \pi_* $ to $ \pi_*|W_u : W_u \to T_{\pi(u)} M $
is an isomorphism.
{}From the above consideration,
it is obvious that there exists a unique metric $ m $ on $ M $
such that $ \pi_*|W_u $ becomes an isometry.
Now $ (M, m) $ becomes a Riemannian manifold
and $ \pi : G \to M $ becomes a Riemannian submersion.
It is also obvious that
\begin{equation}
	l_a^* \, m = m ,
	\label{eqn:2.25}
\end{equation}
namely, $ G $ acts on $ M $ isometrically.
Therefore the metric $ m $ defines the
$ G$-invariant measure $ \mu $ on $ M $.
\subsubsection{Connection}
Note that the decomposition $ T_u G = V_u \oplus W_u $
satisfies the axiom of a connection.
We can define the connection form $ \omega $ for it.
Let $ \h $ be the Lie algebra of the Lie group $ H $,
which is a subalgebra of $ \g $.
$ \g $ is orthogonally decomposed into $ \h \oplus \hp $
with respect to the metric $ \beta $,
where $ \hp $ is the orthogonal complement of $ \h $.
Let $ P : \g \to \h $
be the projection according the above decomposition.
The connection form $ \omega : T G \to \h $ is defined by
\begin{equation}
	\omega (X) := ( P \circ \theta ) (X),
	\qquad
	X \in T G.
	\label{eqn:2.26}
\end{equation}
\par
We close consideration on the intrinsic structures of $ M $ and $ G $.
Next, we proceed to build an additional structure on them, that is to say,
a representation space of the group $ G $.
\subsection{Construction of representation}
Let $ \sigma $ be an $n$-dimensional unitary representation of
the group $ H $.
Define an action of $ h \in H $ on $ G \times \bm{C}^n $ by
\begin{equation}
	\bar{\sigma}(h) : G \times \bm{C}^n \to G \times \bm{C}^n,
	\qquad
	( u, v ) \mapsto ( u h^{-1}, \sigma(h) v ).
	\label{eqn:2.27}
\end{equation}
The associated vector bundle $ \Es = G \times_\sigma \bm{C}^n $
is a quotient space $ G \times \bm{C}^n / H $
in which $ ( u, v ) $ and $ \bar{\sigma}(h) ( u, v ) $ are identified.
The equivalence class of $ ( u, v ) $ is denoted by $ [ u, v ] $.
$ \Es $ is a vector bundle over $ M $
with a fiber $ \bm{C}^n $ and a projection
\begin{equation}
	\pis : \Es \to M,
	\qquad
	[ u, v ] \mapsto \pi(u).
	\label{eqn:2.28}
\end{equation}
We denote a fiber on a point $ x \in M $ by $ \Es|_x = \pis^{-1}(x) $.
Define an inner product fiberwisely by
\begin{equation}
	\bra \, [ u, v ] \, , \, [ u', v' ] \, \ket
	:=
	\bra v , \sigma( u^{-1} u' ) v' \ket,
	\qquad
	[ u, v ], \, [ u', v' ] \in \Es|_x,
	\label{eqn:2.29}
\end{equation}
where the right-hand side is the standard inner product of $ \bm{C}^n $.
It is obvious that
the left-hand side is well-defined as an inner product of $ \Es|_x $.
Define an action of $ a \in G $ on $ \Es $ by
\begin{equation}
	\lambda_a : \Es \to \Es,
	\qquad
	[ u, v ] \mapsto [ a u, v ].
	\label{eqn:2.30}
\end{equation}
Note that $ \lambda_a $ is unitary fiberwisely, $ \Es|_x \to \Es|_{ax} $,
with respect to the inner product (\ref{eqn:2.29}).
Obviously, $ \pis \circ \lambda_a = l_a \circ \pis $.
It should be kept in mind that $ l_a : M \to M $ is isometry
with respect to the metric $ m $ (\ref{eqn:2.25}).
\par
Let $ \psi $ be a square-integrable section of the vector bundle $ \Es $,
which is a differentiable map $ M \to \Es $ such that
$ \pis(\psi(x)) = x $ for $ x \in M $ and
\begin{equation}
	\int_M \bra \psi(x), \psi(x) \ket \, \di \mu (x) < \infty,
	\label{eqn:2.31}
\end{equation}
where $ \bra \: , \: \ket $ refers to
the inner product of $ \Es|_x $, (\ref{eqn:2.29}).
$ \Gs $ denotes the set of the square-integrable sections of $ \Es $.
Define an inner product of $ \Gs $ by
\begin{equation}
	\bra \varphi, \psi \ket
	:=
	\int_M \bra \varphi(x), \psi(x) \ket \, \di \mu (x),
	\qquad
	\varphi, \psi \in \Gs,
	\label{eqn:2.32}
\end{equation}
where $ \bra \: , \: \ket $ in the right-hand side also refers to
the inner product of $ \Es|_x $, (\ref{eqn:2.29}).
The completion of $ \Gs $
with respect to the norm defined by the above inner product
is denoted by the same symbol $ \Gs $
and in what follows completeness is assumed.
Thus $ \Gs $ is a Hilbert space.
We define a map $ \nu : \Gs \times \Gs \to \F(M) $ by
\begin{equation}
	\nu( \varphi, \psi )(x)
	:=
	\bra \varphi(x), \psi(x) \ket,
	\qquad x \in M.
	\label{eqn:2.33}
\end{equation}
Define an action of $ a \in G $ on $ \Gs $ by
\begin{equation}
	\rho(a) : \Gs \to \Gs,
	\qquad
	\psi \mapsto \lambda_a \circ \psi \circ l_a^{-1},
	\label{eqn:2.34}
\end{equation}
that implies
\begin{equation}
	\rho(a) \psi : M \to \Es,
	\qquad
	x \mapsto \lambda_a \, \psi(a^{-1} \, x).
	\label{eqn:2.35}
\end{equation}
Notice that $ \rho $ is a unitary representation of $ G $ on $ \Gs $
and it satisfies the local unitarity condition (\ref{eqn:2.10}).
Hence we have obtained $ ( \Gs, \nu, \rho ) $
for each unitary representation $ \sigma : H \to U(n) $.
Construction of Hamiltonian is postponed
until the section~\ref{sect:Hamiltonian}.
\par
We can express the above argument without use of the vector bundle.
$ \sigma $ is also assumed to be
an $n$-dimensional unitary representation of the group $ H $.
A differentiable map $ \psi^\# : G \to \bm{C}^n $ satisfying
\begin{equation}
	\psi^\# ( u h^{-1} ) = \sigma( h ) \psi^\# ( u ),
	\qquad
	u \in G, \quad h \in H
	\label{eqn:2.36}
\end{equation}
is called a function of $ \# $-type.
Let $ \varphi^\# $ and $ \psi^\# $ be functions of $ \# $-type.
We define a function
$ \nu^\# ( \varphi^\#, \psi^\# ) : M \to \bm{C} $ by
\begin{equation}
	\nu^\# ( \varphi^\#, \psi^\# )(x)
	:=
	\bra \varphi^\# (u), \psi^\# (u) \ket,
	\qquad x \in M, \quad u \in \pi^{-1} (x),
	\label{eqn:2.37}
\end{equation}
where $ \bra \: , \: \ket $ refers to
the standard inner product of $ \bm{C}^n $.
This is well-defined because of the property (\ref{eqn:2.36}).
The function of $ \# $-type is said to be square-integrable if
\begin{equation}
	\int_M \nu^\# ( \psi^\#, \psi^\# )(x) \, \di \mu (x)
	< \infty.
	\label{eqn:2.38}
\end{equation}
$ \Gs^\# $ denotes the set of the square-integrable functions of
$ \# $-type.
Define an inner product of $ \Gs^\# $ by
\begin{equation}
	\bra \varphi^\#, \psi^\# \ket^\#
	:=
	\int_M \nu^\# ( \varphi^\#, \psi^\# )(x) \, \di \mu (x),
	\qquad
	\varphi^\#, \psi^\# \in \Gs^\#.
	\label{eqn:2.39}
\end{equation}
The completion of $ \Gs^\# $
with respect to the norm defined by the above inner product
is also denoted by $ \Gs^\# $.
Therefore $ \Gs^\# $ is a Hilbert space.
Furthermore, define an action of $ a \in G $ on $ \Gs^\# $ by
\begin{equation}
	\rho^\# (a) : \Gs^\# \to \Gs^\#,
	\qquad
	\psi^\# \mapsto \psi^\# \circ L_a^{-1},
	\label{eqn:2.40}
\end{equation}
that implies
\begin{equation}
	\rho^\# (a) \psi^\# : G \to \bm{C}^n,
	\qquad
	u \mapsto \psi^\# (a^{-1} \, u).
	\label{eqn:2.41}
\end{equation}
$ \rho^\# $ is also a unitary representation of $ G $ on $ \Gs^\# $
and satisfies the local unitarity condition (\ref{eqn:2.10}).
Hence we have obtained $ ( \Gs^\#, \nu^\#, \rho^\# ) $,
which also forms quantum mechanics on $ M $.
\par
We shall show equivalence of
$ ( \Gs, \nu, \rho ) $ and $ ( \Gs^\#, \nu^\#, \rho^\# ) $.
It is easily seen that $ \Gs^\# $ is isomorphic to $ \Gs $.
For $ u \in G $, we define a map
\begin{equation}
	\widetilde{u} : \bm{C}^n \to \Es|_{\pi(u)},
	\qquad
	v \mapsto [ u, v ],
	\label{eqn:2.42}
\end{equation}
which is unitary, thus there exists $ ( \widetilde{u} )^{-1} $.
It is obvious that $ \widetilde{uh} = \widetilde{u} \circ \sigma(h) $.
The correspondence between $ \Gs $ and $ \Gs^\# $ is given by
$ \# : \Gs \to \Gs^\#, \, \psi \mapsto \psi^\# $
where $ \psi^\# $ is defined by
\begin{equation}
	\psi^\# (u) =
	( ( \widetilde{u} )^{-1} \circ \psi \circ \pi ) (u).
	\label{eqn:2.43}
\end{equation}
The property (\ref{eqn:2.36}) is verified immediately:
\begin{eqnarray}
	\psi^\# ( u h^{-1} )
	& = &
	( ( \widetilde{ u h^{-1} } )^{-1} \circ \psi \circ \pi )
	( u h^{-1} )
	\nonumber
	\\
	& = &
	( \sigma( h^{-1} )^{-1} \circ ( \widetilde{u} )^{-1}
	\circ \psi \circ \pi ) (u)
	\nonumber
	\\
	& = &
	\sigma(h) \, \psi^\# (u).
	\label{eqn:2.44}
\end{eqnarray}
The inverse $ \#^{-1} : \Gs^\# \to \Gs, \, \psi^\# \mapsto \psi $
is defined by
\begin{equation}
	\psi (x) = [ u, \psi^\# (u) ],
	\qquad x \in M, \quad u \in \pi^{-1} (x).
	\label{eqn:2.45}
\end{equation}
It is obvious that $ \# : \Gs \to \Gs^\# $ is unitary
and $ \nu^\# = \nu \circ \#^{-1}, $
$ \rho^\#(a) = \# \circ \rho(a) \circ \#^{-1} $.
In this sense,
$ ( \Gs, \nu, \rho ) $ and $ ( \Gs^\#, \nu^\#, \rho^\# ) $
are unitary equivalent.
\subsection{Equivalent representations}
In the above argument, a point $ p \in M $ is arbitrarily chosen and fixed.
We shall show that the above construction leads to equivalent result
even if we choose another point $ p' \in M $.
Since the action of $ G $ on $ M $ is transitive,
there exists an element $ k \in G $ such that $ p' = k p $.
$ H $ and $ H' $ denote the isotropy groups of $ p $ and $ p' $ respectively.
They are related by $ H' = k H k^{-1} $.
Two principal fiber bundles, $ ( G, \pi, M, H ) $ and $ ( G, \pi', M, H' ) $
are constructed in the same way.
Define a map $ \kappa_G : G \to G $ by $ u \mapsto u k^{-1} $.
$ \kappa_G $ satisfies the following:
\begin{eqnarray}
	&& \pi = \pi' \circ \kappa_G
	\label{eqn:2.46}
	\\
	&& \kappa_G \circ L_a = L_a \circ \kappa_G,
	\qquad \quad \quad a \in G,
	\label{eqn:2.47}
	\\
	&& \kappa_G \circ R_h = R_{ k h k^{-1}} \circ \kappa_G,
	\qquad h \in H.
	\label{eqn:2.48}
\end{eqnarray}
\par
Let us turn to representations.
Assume that $ \sigma : H \to U(n) $ and $ \sigma' : H' \to U(n) $ are
unitary equivalent representations,
namely, assume that there exists an element $ \epsilon $ of $ U(n) $
such that
$ \sigma'( k h k^{-1} ) = \epsilon \, \sigma( h ) \, \epsilon^\dagger $
$ ( h \in H ) $.
Two representations $ \sigma $ and $ \sigma' $ define
vector bundles $ \pis : \Es \to M $ and $ \pis' : \Es' \to M $
associated to principal fiber bundles
$ \pi : G \to M $ and $ \pi' : G \to M $ respectively.
In what follows, other corresponding objects are indicated by prime.
Define a bundle map $ \kappa_E $ by
\begin{equation}
	\kappa_E : \Es \to \Es' , \qquad
	[ u, v ] \mapsto [ u k^{-1}, \epsilon v ],
	\label{eqn:2.49}
\end{equation}
which is well-defined and fiberwisely unitary.
It is obvious that
\begin{eqnarray}
	&& \pis = \pis' \circ \kappa_E
	\label{eqn:2.50}
	\\
	&& \kappa_E \circ \lambda_a = \lambda_a \circ \kappa_E, \qquad a \in G.
	\label{eqn:2.51}
\end{eqnarray}
The bundle map $ \kappa_E $ induces a map $ \kappa_\Gm $ by
\begin{equation}
	\kappa_\Gm : \Gs \to \Gs' , \qquad
	\psi \mapsto \kappa_E \circ \psi,
	\label{eqn:2.52}
\end{equation}
which is an isometry as a correspondence between Hilbert spaces.
In terms of function of $ \# $-type,
the correspondence between $ \Gs^\# $ and $ \Gs'^\# $ is given by
\begin{equation}
	\kappa_\Gm^\# : \Gs^\# \to \Gs'^\#, \qquad
	\psi^\# \mapsto \epsilon \circ \psi^\# \circ \kappa_G^{-1},
	\label{eqn:2.53}
\end{equation}
which is also well-defined.
It is easily seen that
$ \nu' = \nu \circ \kappa_\Gm^{-1} $ and
$ \rho'(a) = \kappa_\Gm \circ \rho(a) \circ \kappa_\Gm^{-1} $.
Accordingly we conclude that quantum mechanics $ ( \Gs', \nu', \rho' ) $
is equivalent to $ ( \Gs, \nu, \rho ) $.
\subsection{Local expression}
The vector bundle $ \Es $ is locally a direct product
$ U_\alpha \times \bm{C}^n $, where $ U_\alpha $ is an open set of $ M $.
As already stated, the action $ \lambda_a : \Es \to \Es \, ( a \in G ) $
transfers a fiber to a fiber.
Therefore, if $ \lambda_a $ is restricted on $ \pis^{-1} ( U_\alpha ) $,
it can be expressed in terms of linear transformations of $ \bm{C}^n $.
Now we shall show the restricted forms of $ \lambda_a $ and $ \rho(a) $.
By doing it we shall clarify relation
of our formulation to the one of Ohnuki and Kitakado.
\par
Let $ \{ U_\alpha \}_{\alpha \in A} $ be an open covering of $ M $.
Let $ s_\alpha : U_\alpha \to G $ be a local section
of the principal fiber bundle $ ( G, \pi, M, H ) $,
which has a property such that
$ s_\alpha(x) p = x $ for any point $ x \in U_\alpha $ by the definition.
Define a map $ \phi_\alpha (x) : H \to \pi^{-1} (x) $ by
$ h \mapsto s_\alpha (x) \cdot h $.
When $ x \in U_\alpha $ and  $ a^{-1} x \in U_\alpha $ for $ a \in G $,
$ \phi_\alpha ( x )^{-1} \circ L_a \circ \phi_\alpha ( a^{-1} x ) $
is well-defined and it can be identified with an element of $ H $ given by
\begin{equation}
	Q_\alpha ( a, x )
	:= s_\alpha (x)^{-1} \cdot a \cdot s_\alpha ( a^{-1} x ).
	\label{eqn:2.54}
\end{equation}
We call $ Q_\alpha ( a, x ) $ the local expression of
$ L_a $ associated to the local section $ s_\alpha $.
On the other hand, $ s_\alpha(x) $ defines an isomorphism by
\begin{equation}
	\widetilde{s}_\alpha (x) : \bm{C}^n \to \Es|_x, \qquad
	v \mapsto [ s_\alpha(x), v ].
	\label{eqn:2.55}
\end{equation}
The action of $ \lambda_a $ on $ \Es|_{ a^{-1} x } $ is given by
\begin{eqnarray}
	\lambda_a [ s_\alpha( a^{-1} x ), \, v ]
	& = &
	[ a \cdot s_\alpha( a^{-1} x ), \, v ]
	\nonumber \\
	& = &
	[ s_\alpha( x ) \cdot s_\alpha( x )^{-1} \cdot
	a \cdot s_\alpha( a^{-1} x ), \, v ]
	\nonumber \\
	& = &
	[ s_\alpha( x ) \cdot Q_\alpha( a, x ), \, v ]
	\nonumber \\
	& = &
	[ s_\alpha( x ), \, \sigma( Q_\alpha( a, x ) ) v ].
	\label{eqn:2.56}
\end{eqnarray}
Thus we have seen that
\begin{equation}
	\widetilde{s}_\alpha ( x )^{-1} \circ \lambda_a \circ
	\widetilde{s}_\alpha ( a^{-1} x )
	= \sigma( Q_\alpha( a, x ) ).
	\label{eqn:2.57}
\end{equation}
The above equation is called the local expression of
$ \lambda_a $ over $ U_\alpha $.
\par
Moreover, the local section $ s_\alpha : U_\alpha \to G $
gives also the section $ \psi : M \to \Es $ a local expression.
The local expression of $ \psi $ over $ U_\alpha $ is a map
$ \psi_\alpha : U_\alpha \to \bm{C}^n $ defined by
$ \psi_\alpha := \psi^\# \circ s_\alpha $.
It is rewritten as
\begin{equation}
	\psi_\alpha : U_\alpha \to \bm{C}^n, \qquad
	x \mapsto ( \widetilde{s}_\alpha (x)^{-1} \circ \psi ) (x)
	\label{eqn:2.58}
\end{equation}
When $ x \in U_\alpha $ and  $ a^{-1} x \in U_\alpha $,
referring to (\ref{eqn:2.35}),
the local expression of $ \rho(a) \psi $ is calculated as
\begin{eqnarray}
	( \rho(a) \psi )_\alpha (x)
	& = &
	( \widetilde{s}_\alpha (x)^{-1} \circ \rho(a) \psi ) (x)
	\\ \nonumber
	& = &
	( \widetilde{s}_\alpha (x)^{-1} \circ \lambda_a \circ \psi )
	( a^{-1} x)
	\\ \nonumber
	& = &
	( \widetilde{s}_\alpha (x)^{-1} \circ \lambda_a \circ
	\widetilde{s}_\alpha ( a^{-1} x ) ) \circ
	( \widetilde{s}_\alpha ( a^{-1} x )^{-1} \circ \psi )
	( a^{-1} x),
	\label{eqn:2.59a}
\end{eqnarray}
with which using (\ref{eqn:2.57}) and (\ref{eqn:2.58}), we obtain
\begin{equation}
	( \rho(a) \psi )_\alpha (x)
	= \sigma( Q_\alpha ( a, x ) ) \, \psi_\alpha (a^{-1} x).
	\label{eqn:2.59}
\end{equation}
In the context of Ohnuki and Kitakado~\cite{OK1}, \cite{OK2},
$ \sigma( Q_\alpha ( a, x ) ) $ is called the Wigner rotation.
\par
We shall add the transformation rule of the local expressions.
If $ U_\alpha \cap U_\beta \ne \emptyset $,
the transition function $ t_{\alpha \beta} $ is defined by
\begin{equation}
	t_{\alpha \beta} : U_\alpha \cap U_\beta \to H, \qquad
	x \mapsto s_\alpha(x)^{-1} \cdot s_\beta(x),
	\label{eqn:2.60}
\end{equation}
which implies that
$ s_\beta (x) = s_\alpha (x) \, t_{ \alpha \beta } (x) $.
We have already defined the local expression of $ \psi $
by $ \psi_\alpha := \psi^\# \circ s_\alpha $.
Therefore, referring to (\ref{eqn:2.36}),
the local expression of $ \psi $ is transformed as
\begin{eqnarray}
	\psi_\beta (x)
	& = & \sigma( t_{\alpha \beta} (x) )^{-1} \, \psi_\alpha (x)
	\nonumber \\
	& = & \sigma( t_{\beta \alpha} (x) ) \, \psi_\alpha (x).
	\label{eqn:2.62}
\end{eqnarray}
\subsection{Hamiltonian}\label{sect:Hamiltonian}
Here we discuss construction of the Hamiltonian.
We assume that the time evolution of a state vector
$ \psi(t) \in \Gs $ obeys
the Schr\"odinger equation
\begin{equation}
	\k \frac{\di \psi}{\di t} = H \psi
	\label{eqn:2.63}
\end{equation}
as in the ordinary quantum mechanics.
We may introduce various Hamiltonians
because restriction on them is only that they must be self-adjoint.
In this paper we construct a Hamiltonian by the Casimir operator
of the representation $ \rho $.
\par
The unitary representation $ \rho $ of the Lie group $ G $
induces a representation of the Lie algebra $ \g $
by differentiation, that is
$ \rho( A ) := ( \di / \di \xi ) \rho( \exp ( \xi A ) )|_{ \xi=0 } $
for $ A \in \g $.
It is obvious that $ \rho(A)^\dagger = - \rho(A) $.
Let $ \{ X_1, \cdots , X_f \} $ be an orthonormal basis of $ \g $
with respect to the metric $ \beta $.
We define a Hamiltonian $ H $ by
\begin{equation}
	H := \frac 12 \, \sum_{i=1}^f \rho( X_i )^\dagger \, \rho( X_i ),
	\label{eqn:2.64}
\end{equation}
which is hermitian, non-negative and $ G$-invariant (\ref{eqn:2.11}).
\par
Next we shall show that the Hamiltonian acting on $ \Gs^\# $
can be expressed by the Laplacian of $ G $.
For preparation, we review the Laplacian of $ G $ briefly.
Let $ A $ be an element of $ \g = T_e G $.
The left-invariant vector field $ A^L $ and
the right-invariant vector field $ A^R $ over $ G $ associated to $ A $
are defined by
\begin{equation}
	A^L ( u ) := L_{u*} \, A , \qquad
	A^R ( u ) := R_{u*} \, A , \qquad
	u \in G ,
	\label{eqn:2.65}
\end{equation}
respectively.
Since the metric $ g $ of $ G $ is both-invariant,
both of $ \{ X_1^L, \cdots , X_f^L \} $ and $ \{ X_1^R, \cdots , X_f^R \} $
are orthonormal frame fields over $ G $.
Moreover it can be verified that integral curves of $ A^L $ and $ A^R $
are geodesics.
Therefore the Laplacian of $ G $ can be written as
\begin{equation}
	\Delta_G
	= \sum_{i=1}^f ( X_i^L )^2
	= \sum_{i=1}^f ( X_i^R )^2 ,
	\label{eqn:2.66}
\end{equation}
where each term is understood as an operator acting on $ C^\infty ( G ) $.
\par
Differentiating (\ref{eqn:2.41}),
we obtain the representation of $ A \in \g $ on $ \Gs^\# $,
which is given by
\begin{eqnarray}
	\rho^\# ( A ) \psi^\# ( u )
	& = &
	\frac{ \di }{ \di \xi } \, \rho^\# ( \e^{ \xi A } )
	\psi^\# ( u ) \Bigr|_{ \xi=0 }
	\nonumber
	\\
	& = &
	\frac{ \di }{ \di \xi } \, \psi^\# ( \e^{ - \xi A } \cdot u )
	\Bigr|_{ \xi=0 }
	\nonumber
	\\
	& = &
	- A^R \psi^\# ( u ).
	\label{eqn:2.67}
\end{eqnarray}
In the last line, $ A^R $ is understood as a differential operator.
As a consequence,
the representation of the Hamiltonian $ H $ (\ref{eqn:2.64})
on $ \Gs^\# $ is given by
\begin{eqnarray}
	H \psi^\# ( u )
	& = &
	- \frac 12 \, \sum_{i=1}^f ( \rho^\# ( X_i ) )^2 \,
	\psi^\# ( u )
	\nonumber \\
	& = &
	- \frac 12 \, \sum_{i=1}^f ( X_i^R )^2 \, \psi^\# ( u )
	\nonumber \\
	& = &
	- \frac 12 \, \sum_{i=1}^f ( X_i^L )^2 \, \psi^\# ( u )
	\nonumber \\
	& = &
	- \frac 12 \, \Delta_G \, \psi^\# ( u ).
	\label{eqn:2.68}
\end{eqnarray}
$ \g $ has been orthogonally decomposed into
$ \h \oplus \hp $.
Let $ \{ S_1, \cdots , S_p \} $ and $ \{ T_1, \cdots , T_q \} $ be
orthonormal basis of $ \h $ and $ \hp $ respectively.
Because of (\ref{eqn:2.36}),
the action of $ S_i^L $ on $ \psi^\# $ yields
\begin{eqnarray}
	S_i^L \psi^\# ( u )
	& = &
	\frac{\di}{\di \xi} \psi^\# ( u \cdot \e^{ \xi S_i } )
	\Bigr|_{ \xi=0 }
	\nonumber \\
	& = &
	\frac{\di}{\di \xi} \sigma( \e^{ - \xi S_i } ) \psi^\# ( u )
	\Bigr|_{ \xi=0 }
	\nonumber \\
	& = &
	- \sigma( S_i ) \psi^\# ( u ),
	\label{eqn:2.69}
\end{eqnarray}
where the representation of the Lie algebra $ \h $
induced from the representation $ \sigma $ of the Lie group $ H $
is also denoted by $ \sigma $.
We define the Casimir operator $ \cs : \bm{C}^n \to \bm{C}^n $
of $ \sigma $ by
\begin{equation}
	\cs := \sum_{i=1}^p \sigma( S_i )^2 .
	\label{eqn:2.70}
\end{equation}
Furthermore, if we define
\begin{equation}
	\Delta^\# \psi^\# := \sum_{j=1}^q ( T_j^L )^2 \, \psi^\#,
	\label{eqn:2.71}
\end{equation}
we may write $ \Delta_G \psi^\# = \cs \psi^\# + \Delta^\# \psi^\# $.
As readily seen, $ \sigma(h) \cdot \cs \cdot \sigma(h)^{-1} = \cs $
for arbitrary $ h \in H $.
Thus $ ( \cs \psi^\# ) ( u h^{-1} ) = \sigma( h ) ( \cs \psi^\# ) ( u ) $.
On the other hand,
$ ( \Delta_G \psi^\# ) ( u h^{-1} ) = \sigma( h ) ( \Delta_G \psi^\# ) ( u ) $.
Accordingly, both of $ \cs $ and $ \Delta^\# $ are well-defined
operators on $ \Gs^\# $.
Hence $ \Delta := \#^{-1} \circ \Delta^\# \circ \# $ is also
a well-defined operator on $ \Gs $.
Thus we conclude that the Hamiltonian $ H $ is represented
by $ - \frac12 ( \Delta + \cs ) $ on $ \Gs $,
and by $ - \frac12 \Delta_G = - \frac12 ( \Delta^\# + \cs ) $ on $ \Gs^\# $.
\par
If $ \sigma $ is a trivial representation, $ \cs = 0 $
and $ \Delta $ is equal to the Laplacian $ \Delta_M $ of $ M $
as shown in what follows.
If $ \sigma $ is a non-trivial representation,
$ \Delta $ is expressed in terms of covariant derivative
defined by the connection $ \omega $, that is also shown below.
For this purpose, we shall examine the local expression of $ \Delta $
associated to the local section $ s_\alpha : U_\alpha \to G $.
By the definition (\ref{eqn:2.71}),
\begin{equation}
	\Delta^\# \psi^\# ( u )
	= \sum_{j=1}^q
	\frac{ \di^2 }{ \di \xi^2 }
	\psi^\# ( u \cdot \e^{ \xi T_j } ) \Bigr|_{ \xi=0 },
	\label{eqn:2.72}
\end{equation}
thus its pullback by $ s_\alpha $ defines
the local expression of $ \Delta \psi $ as
\begin{equation}
	( \Delta \psi)_\alpha ( x )
	:= ( \Delta^\# \psi^\# ) ( s_\alpha ( x ) )
	= \sum_{j=1}^q
	\frac{ \di^2 }{ \di \xi^2 }
	\psi^\# ( s_\alpha( x )
	\cdot \e^{ \xi T_j } ) \Bigr|_{ \xi=0 }.
	\label{eqn:2.73}
\end{equation}
Now we put
\begin{eqnarray}
	&&
	\widetilde{\gamma}_j ( \xi )
	:= s_\alpha( x ) \cdot \e^{ \xi T_j } ,
	\\
	&&
	\gamma_j ( \xi ) := \pi( \widetilde{\gamma}_j ( \xi ) )
	= s_\alpha( x ) \cdot \e^{ \xi T_j } \cdot p .
	\label{eqn:2.74}
\end{eqnarray}
$ \widetilde{\gamma}_j $ and $ \gamma_j $ are curves on $ G $ and $ M $
respectively.
We denote tangent vectors of $ \widetilde{\gamma}_j $ and $ \gamma_j $
by $ \widetilde{Y}_j ( \xi ) $ and $ Y_j ( \xi ) $ respectively.
Substitution of
$ \widetilde{Y}_j ( \xi ) =  \widetilde{\gamma}_j ( \xi ) \cdot T_j $
into the Maurer-Cartan form $ \theta $ (\ref{eqn:2.15}) yields
\begin{equation}
	\theta ( \widetilde{Y}_j ( \xi ) )
	= ( \widetilde{\gamma}_j ( \xi ) )^{-1} \, \widetilde{Y}_j ( \xi )
	= T_j \in \hp ,
	\label{eqn:2.75}
\end{equation}
with which (\ref{eqn:2.26}) gives $ \omega( \widetilde{Y}_j  ( \xi ) ) = 0 $,
namely $ \widetilde{\gamma}_j $ is a horizontal lift of $ \gamma_j $.
Moreover we define
\begin{equation}
	h_j ( \xi ) :=
	( \widetilde{\gamma}_j (\xi) )^{-1} \cdot s_\alpha( \gamma_j (\xi) ),
	\label{eqn:2.76}
\end{equation}
which is a curve in the Lie group $ H $ and satisfies $ h_j (0) = e $.
Using these notations, the last term in (\ref{eqn:2.73}) is rewritten as
\begin{eqnarray}
	\psi^\# ( s_\alpha( x ) \cdot \e^{ \xi T_j } )
	& = &
	\psi^\# ( \widetilde{\gamma}_j (\xi) )
	\nonumber \\
	& = &
	\psi^\# ( s_\alpha(\gamma_j(\xi))
	\cdot s_\alpha(\gamma_j(\xi))^{-1} \cdot
	\widetilde{\gamma}_j (\xi) )
	\nonumber \\
	& = &
	\sigma( h_j (\xi) ) \cdot
	\psi^\# ( s_\alpha( \gamma_j(\xi) ) )
	\nonumber \\
	& = &
	\sigma( h_j (\xi) ) \cdot
	\psi_\alpha ( \gamma_j(\xi) ).
	\label{eqn:2.77}
\end{eqnarray}
Substitution of the tangent vector of the curve
$ s_\alpha( \gamma_j (\xi) ) = \widetilde{\gamma}_j (\xi) \cdot h_j ( \xi ) $
into the Maurer-Cartan form yields
\begin{equation}
	s_\alpha( \gamma_j (\xi) )^{-1} \,
	\frac{\di}{\di \xi} s_\alpha( \gamma_j (\xi) )
	=
	h_j^{-1} T_j h_j + h_j^{-1} \frac{ \di h_j }{ \di \xi },
	\label{eqn:2.78}
\end{equation}
In the right-hand side of (\ref{eqn:2.78}),
the first term is an element of $ \hp $
and the second term is one of $ \h $.
Therefore we obtain
\begin{equation}
	\omega( s_{\alpha *} Y_j (\xi) )
	= h_j(\xi)^{-1} \frac{ \di h_j }{ \di \xi }.
	\label{eqn:2.79}
\end{equation}
Defining a $ \h $-valued 1-form $ \omega_\alpha $ on $ U_\alpha $
by $ \omega_\alpha := \omega \circ s_{\alpha *} $,
(\ref{eqn:2.79}) is rearranged to be
\begin{equation}
	\frac{ \di h_j }{ \di \xi }
	= h_j(\xi) \, \omega_\alpha ( Y_j (\xi) ).
	\label{eqn:2.80}
\end{equation}
Using (\ref{eqn:2.77}) and (\ref{eqn:2.80}), we obtain
\begin{eqnarray}
	\frac{\di}{\di \xi} \,
	\psi^\# ( s_\alpha( x ) \cdot \e^{ \xi T_j } )
	& = &
	\frac{\di}{\di \xi} \,
	\sigma( h_j (\xi) ) \cdot
	\psi_\alpha ( \gamma_j(\xi) )
	\nonumber \\
	& = &
	\sigma( h_j (\xi) )
	\left(
	\frac{\di}{\di \xi} + ( \sigma \circ \omega_\alpha \circ Y_j ) (\xi)
	\right)
	\psi_\alpha ( \gamma_j(\xi) ).
	\label{eqn:2.81}
\end{eqnarray}
Furthermore, since $ h_j(0) = e $, we obtain
\begin{equation}
	\frac{\di^2}{\di \xi^2} \,
	\psi^\# ( s_\alpha( x ) \cdot \e^{ \xi T_j } )
	\Bigr|_{\xi=0}
	=
	\left(
	\frac{\di}{\di \xi}
	+ ( \sigma \circ \omega_\alpha \circ Y_j ) (\xi)
	\right)^2
	\psi_\alpha ( \gamma_j(\xi) )
	\Bigr|_{\xi=0}.
	\label{eqn:2.82}
\end{equation}
Finally we find that (\ref{eqn:2.73}) is given by
\begin{equation}
	( \Delta \psi)_\alpha ( x )
	= \sum_{j=1}^q
	\left(
	\frac{\di}{\di \xi}
	+ ( \sigma \circ \omega_\alpha \circ Y_j ) (\xi)
	\right)^2
	\psi_\alpha ( \gamma_j(\xi) )
	\Bigr|_{\xi=0}.
	\label{eqn:2.83}
\end{equation}
It can be verified that the curves $ \gamma_j \, ( j=1, \cdots , q ) $
are geodesics through $ x $ at $ \xi=0 $
and the set $ Y_j (0) \, ( j=1, \cdots , q ) $
is orthonormal basis of $ T_x M $.
Hence, if $ \sigma $ is a trivial representation, that is $ \sigma( A ) = 0 $
for all $ A \in \h $,
the right-hand side of (\ref{eqn:2.83}) is reduced to the Laplacian
$ \Delta_M $ of $ M $.
When $ \sigma $ is non-trivial,
let $ ( x^1, \cdots , x^q ) $ be a local coordinate system of $ M $
and express the metric $ m $ and the connection form $ \omega_\alpha $
with it as
\begin{eqnarray}
	&&
	m = \sum_{ \mu, \nu = 1 }^q m_{\mu \nu} (x)
	\, \di x^\mu \otimes \di x^\nu,
	\\
	&&
	\omega_\alpha =  \sum_{ i = 1 }^p \sum_{ \mu = 1 }^q
	A_{\alpha \mu}^i (x) \, S_i \otimes \di x^\mu,
	\label{eqn:2.84}
\end{eqnarray}
and define $ | m (x) | := \det( m_{\mu \nu}(x) ) $.
Using these notations, (\ref{eqn:2.83}) is written as
\begin{eqnarray}
	( \Delta \psi)_\alpha ( x )
	& = &
	\sum_{\mu, \nu=1}^q
	\frac 1{ \sqrt{|m|} }
	\left(
	\frac{\partial}{\partial x^\mu}
	+ \sum_{i=1}^p \sigma( S_i ) A_{\alpha \mu}^i
	\right)
	\nonumber \\
	&&
	\qquad \quad
	\times \sqrt{|m|} \, m^{\mu \nu}
	\left(
	\frac{\partial}{\partial x^\nu}
	+ \sum_{k=1}^p \sigma( S_k ) A_{\alpha \nu}^k
	\right)
	\psi_\alpha ( x ).
	\label{eqn:2.85}
\end{eqnarray}
\subsection{Classification}
Our construction of $ ( \Gs, \nu, \rho, H ) $ is characterized by
$ ( \beta, \sigma ) $,
where $ \beta $ is the adjoint-invariant metric of $ \g $
and $ \sigma $ is the unitary representation of the group $ H $.
Hence $ ( \Gs, \nu, \rho, H ) $ constructed with $ ( \beta, \sigma ) $
is denoted by $ \Gm (\beta, \sigma) $.
The problem of classification of quantum mechanics on $ M $
is stated as follows.
\begin{enum1}
\item	Assume that we have two metrics, $ \beta_1 $ and $ \beta_2 $,
	and two representations, $ \sigma_1 $ and $ \sigma_2 $.
	What condition is necessary and sufficient
	to make $ \Gm ( \beta_2, \sigma_2 ) $
	equivalent to $ \Gm ( \beta_1, \sigma_1 ) $?
\item	Assume that we have a quantum mechanics on $ M $,
	$ ( \Gm, \nu, \rho, H ) $ satisfying the axiom (i)-(iv)
	of section~\ref{sect:definition}.
	Does $ ( \beta, \sigma ) $ exist such that
	$ \Gm (\beta, \sigma) $ is equivalent to
	$ ( \Gm, \nu, \rho, H ) $?
\end{enum1}
At the present time, we have not yet found the answer to the above problem.
%
%
\section{Examples}
Having formulated quantum mechanics on a manifold in general form,
let us now turn to examples.
In the following we shall discuss quantum mechanics on
a sphere $ S^n $ in detail,
a torus $ T^n $ and a projective space $ \RP $ in brief.
By examining the case of sphere,
we will clarify relation of our formulation to the one of Ohnuki and Kitakado
again.
\subsection{$ S^n \, ( n \ge 2 ) $}
First we consider the $ n $-dimensional sphere $ S ^n \, ( n \ge 2 ) $.
The group $ G = SO(n+1) $ acts on it transitively.
The isotropy group is $ H = SO(n) $ in this case.
$ S^n $ is assumed to be embedded in $ \bm{R}^{n+1} $,
which offers the Cartesian coordinates $ ( x^1, \cdots , x^{n+1} ) $.
The coordinates of a point on $ S^n $ are constrained as
\begin{equation}
	( x^1 )^2 + \cdots + ( x^{n+1} )^2 = r^2 = \mbox{const}
	\qquad ( r > 0 ).
	\label{eqn:3.1}
\end{equation}
Furthermore, we put $ z := x^{n+1} $.
As the base point $ p $, we take $ p := ( 0, \cdots , 0, r ) $.
Moreover, we define the opposite point $ \bar{p} := ( 0, \cdots , 0, -r ) $.
We introduce an open covering $ \{ U_+ , U_- \} $ of $ S^n $ by
\begin{eqnarray}
	&&
	U_+ := S^n \setminus \bar{p}
	= \{ \, ( x^1, \cdots , x^{n+1} ) \in S^n \,
	| \, x^{n+1} \ne -r \, \},
	\label{eqn:3.2}
	\\
	&&
	U_- := S^n \setminus p
	= \{ \, ( x^1, \cdots , x^{n+1} ) \in S^n \,
	| \, x^{n+1} \ne r \, \}.
	\label{eqn:3.3}
\end{eqnarray}
The isotropy group $ H $ of $ p $ is embedded in $ G $;
the embedding is represented in matrix form as
\begin{equation}
	\left(
	\begin{array}{c|c}
		h & \begin{array}{c} 0 \\ \vdots \\ 0 \end{array} \\
		\hline
		0 \cdots 0 & 1
	\end{array}
	\right)
	\in SO(n+1),
	\qquad h \in SO(n).
	\label{eqn:3.4}
\end{equation}
We define local sections $ s_\alpha : U_\alpha \to G \, ( \alpha = +,- ) $ by
\begin{eqnarray}
	&&
	s_+(x) :=
	\left(
	\begin{array}{rrcrc}
		1\! -\!	\frac{ x^1 x^1 }{ r(r+z) } &
		- 	\frac{ x^1 x^2 }{ r(r+z) } &
		\cdots &
		- 	\frac{ x^1 x^n }{ r(r+z) } &
			\frac{ x^1 }{ r }
		\vspace{4mm} \\
		-	\frac{ x^2 x^1 }{ r(r+z) } &
		1\! -\!	\frac{ x^2 x^2 }{ r(r+z) } &
		\cdots &
		-	\frac{ x^2 x^n }{ r(r+z) } &
			\frac{ x^2 }{ r }
		\\
		\vdots & \vdots & \ddots
		& \vdots & \vdots \\
		- 	\frac{ x^n x^1 }{ r(r+z) } &
		-	\frac{ x^n x^2 }{ r(r+z) } &
		\cdots &
		1\! -\!	\frac{ x^n x^n }{ r(r+z) } &
			\frac{ x^n }{ r }
		\vspace{4mm} \\
		- \frac{ x^1 }{ r } &
		- \frac{ x^2 }{ r } &
		\cdots &
		- \frac{ x^n }{ r } &
		\frac{ z }{ r }
	\end{array}
	\right),
	\label{eqn:3.5}
	\\
	&& \qquad
	\nonumber \\
	&&
	s_-(x) :=
	\left(
	\begin{array}{rrcrc}
		-1\!+\!	\frac{ x^1 x^1 }{ r(r-z) } &
		    - 	\frac{ x^1 x^2 }{ r(r-z) } &
		\cdots &
		    - 	\frac{ x^1 x^n }{ r(r-z) } &
			\frac{ x^1 }{ r }
		\vspace{4mm} \\
		     	\frac{ x^2 x^1 }{ r(r-z) } &
		1\! -	\frac{ x^2 x^2 }{ r(r-z) } &
		\cdots &
		    - 	\frac{ x^2 x^n }{ r(r-z) } &
		\frac{ x^2 }{ r }
		\\
		\vdots & \vdots & \ddots
		& \vdots & \vdots \\
		   	\frac{ x^n x^1 }{ r(r-z) } &
		    - 	\frac{ x^n x^2 }{ r(r-z) } &
		\cdots &
		1\! -\!	\frac{ x^n x^n }{ r(r-z) } &
		\frac{ x^n }{ r }
		\vspace{4mm} \\
		- \frac{ x^1 }{ r } &
		\frac{ x^2 }{ r } &
		\cdots &
		\frac{ x^n }{ r } &
		\frac{ z }{ r }
	\end{array}
	\right).
	\label{eqn:3.6}
\end{eqnarray}
In what follows, matrix elements of $ s_\alpha $ are denoted by
$ (s_\alpha)_{\mu \nu} \, ( \mu, \nu = 1, \cdots , n+1 ) $.
Matrix elements of the transition function
$ t_{+-} (x) := ( s_+ (x) )^{-1} \cdot s_- (x) \in SO(n)
\, ( x \in U_+ \cap U_- ) $
is given by
\begin{eqnarray}
	&&
	( t_{+-} (x) )_{ i1 } =
	- \delta_{i1} + \frac{ 2 x^i x^1 }{ r^2 - z^2 },
	\nonumber \\
	&&
	( t_{+-} (x) )_{ ij } =
	\delta_{ij} - \frac{ 2 x^i x^j }{ r^2 - z^2 }
	\qquad
	( i = 1, \cdots, n; \, j = 2, \cdots n ).
	\label{eqn:3.7}
\end{eqnarray}
\par
Let $ \g = \so(n+1) $ and $ \h = \so(n) $
be the Lie algebra of $ SO(n+1) $ and $ SO(n) $
respectively.
Define the adjoint-invariant metric $ \beta $ of $ \so(n+1) $
by
\begin{equation}
	\beta( A, B ) := - \frac12 \mbox{tr} ( A B ),
	\qquad A, B \in \so(n+1).
	\label{eqn:3.8}
\end{equation}
The orthonormal basis of $ \h $ and $ \hp $ are denoted by
$ \{ S_{ij} \} \, ( 1 \le i < j \le n ) $ and
$ \{ T_i \} \, ( 1 \le i \le n ) $ respectively,
whose matrix elements are defined by
\begin{eqnarray}
	&&
	( S_{ij} )_{ \mu \nu } :=
	\delta_{ i \mu } \, \delta_{ j \nu }
	- \delta_{ j \mu } \, \delta_{ i \nu },
	\label{eqn:3.9}
	\\
	&&
	( T_i )_{ \mu \nu } :=
	\delta_{ i \mu } \, \delta_{ n+1, \nu }
	- \delta_{ n+1, \mu } \, \delta_{ i, \nu },
	\qquad ( \mu, \nu = 1, \cdots , n+1 ).
	\label{eqn:3.10}
\end{eqnarray}
The pullback of the Maurer-Cartan form $ \theta $ by $ s_\alpha $ is denoted
by $ \theta_\alpha = s_\alpha^{-1} \, \di s_\alpha $.
Matrix elements of $ \theta_+ $ can be calculated straightforwardly
and we obtain
\begin{eqnarray}
	&&
	( \theta_+ )_{ij} =
	\frac1{ r ( r + z ) } \, ( x^i \di x^j - x^j \di x^i ) ,
	\nonumber \\
	&&
	( \theta_+ )_{i,n+1} =
	\frac1r \, \di x^i +
	\frac{ x^i }{ r z ( r + z ) } \sum_{k=1}^n x^k \di x^k ,
	\nonumber \\
	&&
	( \theta_+ )_{n+1,i} = - ( \theta_+ )_{i,n+1} .
	\label{eqn:3.11}
\end{eqnarray}
Having obtained $ \theta_+ $, the metric $ m $ of $ M = S^n $ is calculated as
\begin{eqnarray}
	m
	& = &
	\sum_{i=1}^n \beta( \theta_+ , T_i ) \otimes \beta( \theta_+ , T_i )
	\nonumber \\
	& = &
	\sum_{i=1}^n ( \theta_+ )_{i,n+1} \otimes ( \theta_+ )_{i,n+1}
	\nonumber \\
	& = &
	\frac1{ r^2 }
	\left(
	\sum_{i=1}^n \di x^i \otimes \di x^i + \di z \otimes \di z
	\right) ,
	\label{eqn:3.12}
\end{eqnarray}
which is identified with the standard metric of $ S^n $
except the normalization factor.
Following the definition (\ref{eqn:2.26}),
the pullback of the connection form $ \omega $ by $ s_+ $ is also calculated as
\begin{eqnarray}
	\omega_+
	& = &
	( s_+ )^* \omega
	\nonumber \\
	& = &
	\sum_{ 1 \le i < j \le n } \beta( \theta_+, S_{ij} ) \otimes S_{ij}
	\nonumber \\
	& = &
	\sum_{ 1 \le i < j \le n } ( \theta_+ )_{ij} \otimes S_{ij}
	\nonumber \\
	& = &
	\sum_{ 1 \le i , j \le n }
	\frac1{ r ( r + z ) } \, x^i \di x^j \otimes S_{ij},
	\label{eqn:3.13}
\end{eqnarray}
which coincides with the gauge potential
found by Ohnuki and Kitakado~\cite{OK2}.
A calculation of the curvature form
$ \Om := \di \omega + \omega \wedge \omega $ results in
\begin{eqnarray}
	\Omega_+
	& = &
	\sum_{ 1 \le i < j \le n }
	\biggl[
		\, \frac{z}{r^3} \, \di x^i \wedge \di x^j
		\nonumber \\
	&& \hspace{12mm}
		+ \sum_{ 1 \le k \le n }
		\frac{ x^k }{ r^3 ( r+z ) } \,
		( x^i \di x^j \wedge \di x^k
		+ x^j \di x^k \wedge \di x^i
		+ x^k \di x^i \wedge \di x^j )
		\nonumber \\
	&& \hspace{12mm}
		+ \frac1{r^3} \, ( x^i \di x^j - x^j \di x^i ) \wedge \di z
	\biggr]
	\otimes S_{ij},
	\label{eqn:3.14}
\end{eqnarray}
which also coincides with the field strength found by them.
\par
Of course, we can calculate the pullbacks of
$ \theta, \, \omega $ and $ \Om $ by $ s_- $ in a similar way.
Here we show only $ \theta_- $ and $ \omega_- $:
\begin{eqnarray}
	&&
	( \theta_- )_{1j} =
	- \frac1{ r ( r - z ) } \, ( x^1 \di x^j - x^j \di x^1 ) ,
	\nonumber \\
	&&
	( \theta_- )_{1,n+1} =
	- \frac1r \, \di x^1
	+ \frac{ x^1 }{ r z ( r - z ) } \sum_{k=1}^n x^k \di x^k ,
	\nonumber \\
	&&
	( \theta_- )_{ij} =
	\frac1{ r ( r - z ) } \, ( x^i \di x^j - x^j \di x^i ) ,
	\nonumber \\
	&&
	( \theta_- )_{i,n+1} =
	\frac1r \, \di x^i
	- \frac{ x^i }{ r z ( r - z ) } \sum_{k=1}^n x^k \di x^k
	\qquad
	( i , j = 2, \cdots n ),
	\nonumber \\
	&&
	( \theta_- )_{ \nu \mu } = - ( \theta_- )_{ \mu \nu } ,
	\label{eqn:3.15}
	\\
	&&
	\omega_-
	=
	\frac1{ r ( r - z ) }
	\Bigl(
	- \! \sum_{ 2 \le j \le n }
	(x^1 \di x^j - x^j \di x^1 ) \otimes S_{1j}
	+ \! \sum_{ 2 \le i , j \le n } \! x^i \di x^j \otimes S_{ij}
	\Bigr).
	\label{eqn:3.16}
\end{eqnarray}
\par
In the case of $ S^n \, ( n \ge 2 ) $,
our method to construct the representation space is essentially the same
as the one of Ohnuki and Kitakado.
They noticed that the gauge potential associated to the little group
is inevitably introduced in their formalism.
They also noticed that the field strength exhibits the monopole-like structure
which has a singularity at the center of the sphere.
Their gauge potential corresponds to our connection form $ \omega $,
which reflects the geometry of the principal fiber bundle
$ ( G, \pi, M, H ) $.
The singularity manifests non-trivial topology of the principal fiber bundle.
\par
We would like to give physical interpretation
to our formulation of quantum mechanics.
However, consideration on it is postponed
until the section~\ref{sect:interpretation}.
\subsection{$ S^2 $}
In the previous example,
we took $ SO(n+1) $ as the transformation group $ G $
acting on $ S^n \, ( n \ge 2 ) $ transitively.
However, we can also take the spinor group $ Spin(n+1) $ as $ G $.
$ Spin(n+1) $ is the universal covering group of $ SO(n+1) $;
$ Spin(n+1) $ covers $ SO(n+1) $ twofold.
If we construct a representation of $ Spin(n+1) $,
we may have a double-valued representation of $ SO(n+1) $.
\par
Actually, what Ohnuki and Kitakado have constructed are representations of
not $ SO(n+1) $ but $ Spin(n+1) $.
They began with the Lie algebra $ \so(n+1) $
instead of the Lie group $ SO(n+1) $ itself.
It is well known that the representation of
the Lie algebra $ \so(n+1) $ is equivalent to
the representation of the universal covering group $ Spin(n+1) $.
\par
Here we take $ Spin(3) \cong SU(2) $ as $ G $ acting on $ S^2 $.
The isotropy group $ H $ is isomorphic to $ U(1) $.
In what follows, we will show explicitly
that we obtain
both of single-valued and double-valued representations of $ SO(3) $.
\par
We define coordinates $ ( \theta, \phi, \psi ) $ of $ u \in SU(2) $ by
\begin{eqnarray}
	&&
	u ( \theta, \phi, \psi ) =
	\e^{ - \sk \phi \,\sigma_3 / 2 }   \,
	\e^{ - \sk \theta \sigma_2 / 2 } \,
	\e^{ - \sk \psi   \sigma_3 / 2 }
	\nonumber \\
	&&
	\qquad
	( 0 \le \theta \le \pi, \;
	0 \le \phi < 2 \pi, \;
	0 \le \psi < 4 \pi ),
	\label{eqn:3.17}
\end{eqnarray}
where $ \sigma_i ( i =1,2,3 ) $ are the Pauli matrices.
Let $ x = ( x^1, x^2, x^3 ) $ be a point of $ S^2 $ constrained as
$ \sum_{i=1}^3 ( x^i )^2 = 1 $.
We put $ \bar{x} := \sum_{i=1}^3 x^i \sigma_i $
and define the action of $ u $ on $ x $ by
$ \bar{x} \mapsto u \bar{x} u^\dagger $.
In this action,
$ ( \theta, \phi, \psi ) $ are identified with the Euler angles.
We take $ p = ( 0,0,1 ) $ as the base point of $ S^2 $.
The isotropy group of $ p $ is
\begin{equation}
	H = \{ \, h( \xi ) = \e^{ - \sk \xi \sigma_3 / 2 }
	\; ( 0 \le \xi < 4 \pi ) \, \}.
	\label{eqn:3.18}
\end{equation}
By the correspondence $ h ( \xi ) \mapsto \e^{ - \sk \xi / 2 } $,
$ H $ is isomorphic to $ U(1) $.
\par
Notice that $ S^2 $ is isometric to
the 1-dimensional complex projective space $ \CP $.
The above argument can be rephrased
in terms of the action of $ SU(2) $ on $ \CP $.
Let $ [ z^0, z^1 ] $ be the homogeneous coordinates of $ \CP $
and take a representative such that $ | z^0 |^2 + | z^1 |^2 = 1 $.
If we put $ z := {z^0 \choose z^1} $,
the correspondence between $ S^2 $ and $ \CP $ is given by
$ x^i = z^\dagger \sigma_i z $ or $ \bar{x} = 2 z z^\dagger - 1 $,
and the transformation $ \bar{x} \mapsto u \bar{x} u^\dagger $
is equivalent to
the action of $ SU(2) $ on $ \CP $ defined by $ z \mapsto u z $.
Obviously, the base point $ p = ( 0,0,1 ) \in S^2 $
corresponds to $ {1 \choose 0} \in \CP $.
\par
Now we define the adjoint-invariant metric $ \beta $ of
$ \su $
by
\begin{equation}
	\beta( A, B ) := - 2 \mbox{tr} ( A B ),
	\qquad A, B \in \su.
	\label{eqn:3.19}
\end{equation}
For this normalization,
$ \{ - \k \, \sigma_i / 2 \}_{i = 1, 2, 3} $
is an orthonormal basis.
Referring to (\ref{eqn:2.18}),
the metric $ g $ of $ SU(2) $ is calculated to be
\begin{equation}
	g =
	\di \theta^2 +
	\sin^2 \theta \, \di \phi^2 +
	( \di \psi + \cos \theta \, \di \phi )^2.
	\label{eqn:3.20}
\end{equation}
We can also calculate the Laplacian $ \Delta_G $ for this metric to obtain
\begin{equation}
	\Delta_G =
	\frac1{\sin \theta} \frac{\partial}{\partial \theta}
	\biggl( \sin \theta \frac{\partial}{\partial \theta} \biggr)
	+
	\frac1{\sin^2 \theta}
	\biggl( \frac{\partial}{\partial \phi}
	- \cos \theta \frac{\partial}{\partial \psi} \biggr)^2
	+
	\frac{ \partial^2 }{ \partial \psi^2 }.
	\label{eqn:3.21}
\end{equation}
\par
Here we shall consider only 1-dimensional unitary representations of
$ H = U(1) $,
because any unitary representation of $ U(1) $
is reducible to 1-dimensional representations.
The 1-dimensional unitary representation of $ U(1) $ is characterized
by an integer $ n $ and is defined by
\begin{equation}
	\sigma_n : U(1) \to U(1),
	\qquad
	\e^{ - \sk \xi / 2 } \mapsto \e^{ - \sk n \xi / 2 }.
	\label{eqn:3.22}
\end{equation}
With $ \sigma_n $, we can construct the representation space
$ \Gn $ and $ \Gn^\# $ according to the general argument of
section~\ref{sect:Def}.
The right-translation of $ u ( \theta, \phi, \psi ) $ by $ h( \xi )^{-1} $
gives
$ u ( \theta, \phi, \psi ) \, h(\xi)^{-1} = u ( \theta, \phi, \psi - \xi )$.
Therefore, referring to (\ref{eqn:2.36}),
the function of $ \#$-type $ f^\# : SU(2) \to \bm{C} $ satisfies
\begin{equation}
	f^\# ( \theta, \phi, \psi - \xi )
	=
	\e^{ - \sk n \xi / 2 } \, f^\# ( \theta, \phi, \psi ),
	\label{eqn:3.23}
\end{equation}
from which we deduce that
\begin{equation}
	\frac{\partial}{\partial \psi}
	f^\# ( \theta, \phi, \psi )
	=
	\k \frac{n}{2} \, f^\# ( \theta, \phi, \psi ).
	\label{eqn:3.24}
\end{equation}
Hence operation of the Laplacian (\ref{eqn:3.21}) on $ f^\# $ gives
\begin{eqnarray}
	\Delta_G \, f^\#
	& = &
	\left[
	\frac1{\sin \theta} \frac{\partial}{\partial \theta}
	\Bigl( \sin \theta \frac{\partial}{\partial \theta} \Bigr)
	+
	\frac1{\sin^2 \theta}
	\Bigl( \frac{\partial}{\partial \phi}
	- \k \frac{n}{2} \cos \theta \Bigr)^2
	- \frac14 n^2
	\right]
	f^\#
	\nonumber \\
	& = &
	( \Delta^\# + c_n ) f^\#,
	\label{eqn:3.25}
\end{eqnarray}
where we define $ c_n := - n^2 / 4 $, which is the Casimir of $ \sigma_n $.
$ - \frac12 \Delta^\# $
is identical to the Hamiltonian for a particle on $ S^2 $
influenced by the vector potential $ A = - ( n / 2 ) \cos \theta \, \di \phi $
of the monopole with quantum number $ n $.
\par
These results again coincide with that of Ohnuki and Kitakado.
The eigenfunctions of $ \Delta_G $ are completely known
and are given in terms of Jacobi polynomials~\cite{Ohnuki}.
If we put $ S := n / 2 $,
the eigenvalues of $ \Delta_G $ are known to be
$ - j(j+1) $ with $ j = |S|, |S|+1, |S|+2, \cdots $.
If we regard $ f^\# \mapsto \rho^\#(a) f^\# $ as a representation of $ SO(3) $
and if $ n $ is odd,
it is to be a double-valued representation because of (\ref{eqn:3.23}).
More explicitly, if we put
$ a = \e^{ - \sk \pi \sigma_3 } = {-1 \; \: 0 \choose 0 \; \: -1} $,
then $ a \bar{x} a^\dagger = \bar{x} $.
However,
$ a^{-1} u ( \theta, \phi, \psi ) = u ( \theta, \phi, \psi - 2 \pi ) $,
thus we deduce
\begin{equation}
	\rho^\# (a) f^\# (u) = ( -1 )^n f^\# (u)
	\label{eqn:3.25a}
\end{equation}
from (\ref{eqn:2.41}) and (\ref{eqn:3.23}).
\subsection{$ S^1 $ and $ T^n $}
Consideration on the circle $ S^1 $ has been postponed until now.
The sphere $ S^n $ is isometric to the quotient space $ SO(n+1)/SO(n) $.
However, $ S^1 $ is isometric to the group manifold $ SO(2) $ itself.
If we take $ SO(2) $ as a transformation group $ G $ acting on $ S^1 $,
the construction of quantum mechanics on $ S^1 $ results in
only a trivial representation $ ( \Gm^\circ, \rho^\circ ) $;
the Hilbert space $ \Gm^\circ $ is equivalent to
the space of complex-valued square-integrable functions on $ S^1 $,
which is denoted by $ \Lb_{\, 2}(S^1) $;
the representation $ \rho^\circ $ is given by
$ \rho^\circ (a) \psi(x) = \psi( a^{-1} x ) \, ( a \in SO(2), \, x \in S^1 ) $.
\par
Here we take the additive group $ \bm{R} $ as $ G $ acting on $ S^1 $
to induce multi-valued representations of $ SO(2) $.
Notice that $ \bm{R} $ is the universal covering group of $ SO(2) $.
We define the action of $ a \in \bm{R} $ on
$ x \in S^1 = \{ \e^{ \sk \theta } \, | \, \theta \in \bm{R} \} $
by $ x \mapsto \e^{ \sk a } \cdot x $.
The base point $ p \in S^1 $ can be chosen arbitrarily.
The isotropy group is
\begin{equation}
	H = \{ \, 2 \pi n \, | \, n \in \bm{Z} \, \} \cong \bm{Z}.
	\label{eqn:3.26}
\end{equation}
Notice that $ H $ is a discrete group,
therefore its Lie algebra $ \h $ is $ \{ 0 \} $.
Thus both of the connection form and the curvature form vanish.
Furthermore, since
any unitary representation of $ \bm{Z} $ is reducible to 1-dimensional ones,
it is sufficient to consider an only 1-dimensional one.
The 1-dimensional unitary representation of $ H \cong \bm{Z} $ is characterized
by a parameter $ \alpha \in \bm{R} \, ( \mbox{mod } 1 ) $ and is defined by
\begin{equation}
	\sigma_\alpha : H \to U(1),
	\qquad
	2 \pi n \mapsto \e^{ - \sk \, \alpha \, 2 \pi n }.
	\label{eqn:3.27}
\end{equation}
With $ \sigma_\alpha $, referring (\ref{eqn:2.36}),
the function of $ \#$-type $ f^\# : \bm{R} \to \bm{C} $ satisfies
\begin{equation}
	f^\# ( u - 2 \pi n ) = \e^{ - \sk \, \alpha \, 2 \pi n } \, f^\# ( u ).
	\label{eqn:3.28}
\end{equation}
Of course, referring (\ref{eqn:2.41}),
the representation of $ a \in \bm{R} $ on $ f^\# $ is defined by
\begin{equation}
	\rho^\# ( a ) f^\# ( u ) = f^\# ( u - a ),
	\label{eqn:3.29}
\end{equation}
which also defines the action of $ T = 1 \in \Lieu $ on $ f^\# $ by
\begin{equation}
	\rho^\# ( T ) f^\# ( u )
	=
	\frac{ \di }{ \di \xi } f^\# ( u - \xi )
	\Bigr|_{ \xi = 0 }
	=
	- \frac{ \di }{ \di u } f^\# ( u ),
	\label{eqn:3.30}
\end{equation}
referring (\ref{eqn:2.67}).
\par
If we introduce $ f^\circ : \bm{R} \to \bm{C} $ by
\begin{equation}
	f^\circ ( u ) := \e^{ - \sk \alpha u } \, f^\# ( u ),
	\label{eqn:3.31}
\end{equation}
it is a periodic function, $ f^\circ ( u - 2 \pi n ) = f^\circ ( u ) $.
Here we put $ \Gm^\circ := \Lb_{\, 2}(S^1)$.
Thus (\ref{eqn:3.29}) is equivalent to
\begin{equation}
	\rho^\circ_\alpha ( a ) f^\circ ( u )
	= \e^{ - \sk \alpha \, a } \,f^\circ ( u - a ).
	\label{eqn:3.32}
\end{equation}
Moreover, (\ref{eqn:3.30}) is equivalent to
\begin{equation}
	\rho^\circ_\alpha ( T ) f^\circ ( u )
	=
	- \Bigr( \frac{ \di }{ \di u } + \k \alpha \Bigl) f^\circ ( u ).
	\label{eqn:3.33}
\end{equation}
We have seen that
$ ( \Ga^\#, \rho^\# ) $ and $ ( \Gm^\circ, \rho^\circ_\alpha ) $
offer equivalent representations of $ \bm{R} $.
If we regard $ f^\# $ and $ f^\circ $ as functions on $ S^1 $,
$ f^\# $ is multi-valued and $ f^\circ $ is single-valued.
Moreover,
if we regard $ \rho^\circ_\alpha $ as a representation of $ SO(2) $,
it is multi-valued.
$ ( \Gm^\circ, \rho^\circ_\alpha ) $ is identical to
what Ohnuki and Kitakado~\cite{OK2} have constructed.
It is a trivial thing to generalize the above argument
to the case of $ n$-dimensional torus $ T^n = S^1 \times \cdots \times S^1 $.
In that case,
there exists an inequivalent representation for each value of parameters
$ ( \alpha_1, \cdots , \alpha_n ) \in \bm{R}^n / \bm{Z}^n $.
\subsection{$ RP^n $}
As the final example, we shall glance at quantum mechanics on
the real $ n$-dimensional projective space $ \RP $.
$ \RP $ is a non-orientable manifold,
however, orientability is not a matter for our formalism.
\par
$ \RP $ is diffeomorphic to some manifolds, for example,
\begin{eqnarray}
	\RP
	& \cong & O(n+1) / O(n) \times O(1)
	\nonumber \\
	& \cong & SO(n+1) / S( O(n) \times O(1) )
	\nonumber \\
	& \cong & S^n / \bm{Z}_2 ,
	\label{eqn:3.34}
\end{eqnarray}
where $ O(1) = \{ 1, -1 \} \cong \bm{Z}_2 $;
$ S( O(n) \times O(1) ) $ will be explained later.
In this paper we take $ SO(n+1) $ as a transformation group $ G $.
With the double covering $ S^n \to \RP $
and the embedding $ S^n \to \bm{R}^{n+1} $,
$ \bm{R}^{n+1} $ offers the double-valued Cartesian coordinates
$ ( x^1, \cdots, x^{n+1} ) $ to $ \RP $.
The base point $ p $ is put at $ ( 0, \cdots, 0, r ) $.
The isotropy group $ H $ is
\begin{eqnarray}
	H
	& = & S( O(n) \times O(1) )
	\nonumber \\
	& = &
	\{
	\left(
	\begin{array}{c|c}
		h_1 & \begin{array}{c} 0 \\ \vdots \\ 0 \end{array} \\
		\hline
		0 \cdots 0 & h_2
	\end{array}
	\right)
	| \,
	h_1 \in O(n), \: h_2 \in O(1), \: h_2 \det h_1 = 1
	\, \}.
	\label{eqn:3.35}
\end{eqnarray}
If we define
\begin{equation}
	N := SO(n) \times I_1 =
	\{
	\left(
	\begin{array}{c|c}
		h & \begin{array}{c} 0 \\ \vdots \\ 0 \end{array} \\
		\hline
		0 \cdots 0 & 1
	\end{array}
	\right)
	| \,
	h \in SO(n)
	\, \},
	\label{eqn:3.36}
\end{equation}
$ N $ is a normal subgroup of $ H $, and $ H / N \cong O(1) $,
hence we have an isomorphism
$ H \cong SO(n) \times O(1) $.
Therefore a unitary representation of $ H $ can be decomposed into
tensor products of representations of $ SO(n) $ and $ O(1) $.
\par
There are only two irreducible representations of $ O(1) = \{ 1, -1 \} $.
One is trivial representation $ \zeta_+ $,
another is $ \zeta_- $; they are defined by
\begin{eqnarray}
	&&
	\zeta_+ (1) = \zeta_+ (-1) = 1,
	\label{eqn:3.37}
	\\
	&&
	\zeta_- (1) = 1, \quad \zeta_- (-1) = -1.
	\label{eqn:3.38}
\end{eqnarray}
Given a representation $ \sigma : SO(n) \to U(j) $,
we obtain two representations
$ \sigma_\pm := \sigma \otimes \zeta_\pm : SO(n) \times O(1) \to U(j) $.
For each $ \sigma_\pm $, we can construct quantum mechanics on $ \RP $,
$ ( \Gm_{\sigma \pm}, \rho_\pm, \nu_\pm, H_\pm ) $.
On the other hand, with $ \sigma $,
we can also construct quantum mechanics on $ S^n $, $ ( \Gs, \rho, \nu, H ) $.
It is obvious that $ \Gs $ can be decomposed to
$ \Gm_{\sigma_+} \oplus \Gm_{\sigma_-} $
as representation spaces of $ G = SO(n+1) $.
\par
We have taken $ SO(n+1) $ as $ G $ acting on $ \RP $.
However, we may introduce another group.
For instance, since $ \bm{R}P^3 \cong SO(3) \cong SU(2) / \bm{Z}_2 $,
we can take $ SU(2) $ as $ G $ for $ \bm{R}P^3 $.
Schulman~\cite{Schulman} have discussed quantum mechanics of a rigid body
in 3-dimensional space from view point of path integral.
The configuration space of a rigid body is $ \bm{R}P^3 \cong SO(3) $.
He have found that
integer or half-integer angular momentum appears
according as $ \zeta_+ $ or $ \zeta_- $ is taken
for a representation of $ \bm{Z}_2 $.
His argument is included in our formulation;
we regard $ \bm{R}P^3 $ as $ SO(4) / S( O(3) \times O(1) ) $.
If we take the trivial representation $ \sigma $ of $ SO(3) $,
our formulation reproduces his result.
%
%
\section{Discussions}
\subsection{Summary}
Let us summarize subjects studied in this paper.
We have defined quantum mechanics on a homogeneous space $ M = G / H $
and have presented a method to realize the definition.
The realization is
a unitary representation $ \rho $ of the transformation group $ G $
on a space of vector bundle-valued functions on $ M $.
A representation is characterized by $ ( \beta, \sigma ) $,
where $ \beta $ is an adjoint-invariant metric of the Lie algebra $ \g $
and $ \sigma $ is a unitary representation of the isotropy group $ H $.
When $ H \ne \{ e \} $,
there exist a number of inequivalent realizations.
\par
Furthermore,
we have defined a Hamiltonian by the Casimir operator of $ \rho $.
Expressing the Hamiltonian with a local coordinate,
we have seen that it is the Laplacian
in which a partial derivative is replaced by a covariant derivative
with a gauge field.
The gauge field associated with the group $ H $
is automatically introduced in the expression of the Hamiltonian.
\par
As examples, we have studied quantum mechanics on
a sphere $ S^n $, a torus $ T^n $ and a projective space $ \RP $.
In any case,
it is shown that there are an infinite number of inequivalent realizations;
for $ T^n $, there are uncountably infinite realizations;
for the other manifolds, there are countably infinite ones.
Particularly, for $ S^n \, ( n \ge 2 ) $,
if we take $ Spin(n+1) $ as the transformation group acting on $ S^n $,
our construction of quantum mechanics on $ S^n $ leads to
the same result as the one of Ohnuki and Kitakado~\cite{OK2}.
The gauge field exhibits the monopole-like structure.
\subsection{Interpretation}\label{sect:interpretation}
Here we would like to give physical interpretation to our formalism,
in particular, to the definitions (i)-(iv) in section~\ref{sect:definition}.
\subsubsection{Role of position operator}
To compare our formalism with the ordinary operator formalism,
we shall construct quantum mechanics on a Euclidean space $ M = \bm{R}^n $
by our formalism.
We take $ \bm{R}^n $ itself as the transformation group $ G $ acting on $ M $
translationally by $ G \times M \to M, \: ( a, x ) \mapsto x + a $.
In this case, the isotropy group is trivial, that is, $ H = \{ e \} $.
We take the 1-dimensional representation $ \sigma : H \to U(1) $.
In this case,
the Hilbert space $ \Gs $ is identical to $ \Lb_{\, 2}(\bm{R}^n) $,
the space of complex-valued square-integrable functions on $ \bm{R}^n $.
Moreover, $ \rho $ and $ \nu $ is given by
\begin{eqnarray}
	&&
	\rho(a) \psi(x) = \psi(x-a), \qquad \quad a \in G, \quad x \in M,
	\label{eqn:4.1}
	\\
	&&
	\nu( \varphi, \psi ) (x) = \varphi(x)^* \, \psi(x),
	\qquad \varphi, \psi \in \Gs.
	\label{eqn:4.2}
\end{eqnarray}
\par
On the other hand, the ordinary operator formalism is defined as follows.
Assume that $ \hat{x}^i $ and $ \hat{p}_j \, ( i,j = 1, \cdots n ) $
are self-adjoint operators satisfying the canonical commutation relations:
\begin{equation}
	[ \, \hat{x}^i , \hat{p}_j \, ] = \k \, \delta^{\,i}_{\:j} , \qquad
	[ \, \hat{x}^i , \hat{x}^j \, ] =
	[ \, \hat{p}_i , \hat{p}_j \, ] = 0.
	\label{eqn:4.3}
\end{equation}
$ \hat{x}^i $'s and $ \hat{p}_j $'s generate an algebra $ \A $.
The irreducible representation space $ \Hil $ of $ \, \A $ is unique
in the sense of unitary equivalence class.
If we denote a simultaneous eigenstate of position operators $ \hat{x}^i $ by
$ | x \ket = | x^1, \cdots, x^n \ket $, which satisfies
\begin{eqnarray}
	&&
	\hat{x}^i | x \ket = x^i | x \ket,
	\label{eqn:4.4}
	\\
	&&
	\bra x | x' \ket = \delta^n ( x - x' ),
	\label{eqn:4.5}
	\\
	&&
	\int_{ -\infty}^\infty | x \ket \bra x | \, \di^n x = I,
	\label{eqn:4.6}
\end{eqnarray}
and if we define a correspondence
$ \Hil \to \Gs $ by $ | \psi \ket \mapsto \psi(x) := \bra x | \psi \ket $,
$ \Hil $ is unitary equivalent to $ \Gs = \Lb_{\, 2}(\bm{R}^n) $.
Putting
$ \hat{\rho} ( a ) := \exp( - \k \sum_{j=1}^n a^j \hat{p}_j ) $
for $ a \in G $, it is easily seen that
\begin{equation}
	\bra x | \hat{\rho} (a) | \psi \ket = \bra x - a | \psi \ket,
	\label{eqn:4.7}
\end{equation}
which corresponds to (\ref{eqn:4.1}). Of course,
\begin{equation}
	\bra \varphi | \psi \ket =
 	\int_{ -\infty }^\infty
 	\bra \varphi | x \ket \bra x | \psi \ket \di^n x,
 	\qquad
	| \varphi \ket, | \psi \ket \in \Hil,
	\label{eqn:4.8}
\end{equation}
corresponds to (\ref{eqn:4.2}).
\par
{}From the above observation, we can state roles of
the position operators $ \hat{x}^i $'s and
the momentum operators $ \hat{p}_j $'s as follows.
A set of eigenvalues of $ \hat{x}^i $'s has
one-to-one correspondence with a point of $ M $.
Namely, spectrum of $ \hat{x}^i $'s devotes itself
as a coordinate system of $ M $.
$ \hat{p}_j $'s generate translation on $ M $.
Namely, $ \hat{V} = \sum_{j=1}^n a^j \hat{p}_j $ can be identified with
a vector field $ V = \sum_{j=1}^n a^j \partial_j $ on $ M $
and
$ \hat{\rho}(a) = \exp( - \k \hat{V} ) $ can be identified with
an action of $ a \in G $ on $ M $ by $ x \mapsto x + a $.
\par
In general, we may consider a particle on $ \bm{R}^n $
with an extra degree of freedom such as spin.
In such a case, we denote complete orthonormal system by
$ \{ | x^1, \cdots, x^n, s \ket \} $ with an extra index $ s $.
Furthermore, we introduce projection-valued measures
$ \hat{P} $ and $ \hat{P}^i \, ( i = 1, \cdots, n ) $ by
\begin{eqnarray}
	&&
	\hat{P} ( \di x^1 \cdots \di x^n ) :=
	\sum_s | x^1, \cdots, x^n, s \ket \bra x^1, \cdots, x^n, s |
	\, \di x^1 \cdots \di x^n,
	\label{eqn:4.9}
	\\
	&&
	\hat{P}^i ( \di x^i ) :=
	\nonumber \\
	&&
	\quad
	\di x^i
	\sum_s \int_{ -\infty }^\infty
	\di x^1 \cdots \di x^{i-1} \di x^{i+1} \cdots \di x^n \,
	| x^1, \cdots, x^n, s \ket \bra x^1, \cdots, x^n, s |.
	\label{eqn:4.10}
\end{eqnarray}
These give spectral resolution of $ \hat{x}^i $'s:
\begin{equation}
	\hat{x}^i = \int_{ -\infty }^\infty x^i \, \hat{P}^i ( \di x^i ).
	\label{eqn:4.11}
\end{equation}
\par
Now we shall point out difficulties of use of position operators
in quantum mechanics on a general manifold.
There are two points to be noticed.
\par
The first point is that
a general manifold can not be covered with a single coordinate system.
Since $ \hat{x}^i $'s are self-adjoint and commutative,
their spectrum is real
and their simultaneous eigenstates form a complete orthonormal system.
Therefore,
the spectrum can be used as a coordinate system of $ M = \bm{R}^n $.
The completeness of the simultaneous eigenstates means that
$ M $ must be covered with a single coordinate system.
However, it is impossible to parametrize points on a general manifold
continuously by a single coordinate system.
\par
The second point is much subtler.
It annoys us when there exists an extra degree of freedom such as spin.
The point is that
the spinor wave function cannot become a continuous function,
even if points on the manifold can be parametrized continuously somehow.
For example, in quantum mechanics on $ M=S^n \, ( n \ge 2 ) $,
we have used the Cartesian coordinates $ ( x^1, \cdots, x^{n+1} ) $.
This coordinate system is redundant,
because the radius coordinate is not needed.
However it parametrizes points on $ S^n $ continuously.
Actually, Ohnuki and Kitakado~\cite{OK2} have used
$ ( \hat{x}^1, \cdots, \hat{x}^{n+1} ) $ as position operators.
So far, there is no problem.
The problem lies in the extra degree of freedom.
Assume that we take a non-trivial representation $ \sigma : H=SO(n) \to U(j) $.
Given a local section $ s_\alpha : U_\alpha \to G=SO(n+1) $,
$ \psi \in \Gs $ has a local expression
$ \psi_\alpha = \psi^\# \circ s_\alpha : U_\alpha \to \bm{C}^j $.
It can be written in the Dirac's notation as
\begin{equation}
	\psi_\alpha(x) =
	\left(
	\begin{array}{c}
	\bra x, s=1 | \psi \ket \\
	\vdots \\
	\bra x, s=j | \psi \ket
	\end{array}
	\right),
	\qquad x \in U_\alpha.
	\label{eqn:4.12}
\end{equation}
If and only if $ s_\alpha $ is a global section $ s_M : M \to G $,
$ \psi^{(s)} (x) = \bra x, s | \psi \ket $ becomes
a well-defined continuous function on $ M $ for an arbitrary $ \psi \in \Gs $.
In fact, there is no global section $ s_M $ in this case.
Thus $ \bra x, s | \psi \ket $ cannot be a continuous function over $ M $.
\par
To put it briefly, the first point is a matter of topology of $ M $
and the second is a matter of topology of the fiber bundle
$ ( G, \pi, M, H ) $.
\par
The above difficulties suggests that
we should give up use of position operators in quantum mechanics
on a general manifold
and we should seek for more flexible substitutions for them.
If we observe position of a quantum-mechanical particle moving on a manifold,
the probability distribution may be more directly observable
than the coordinates.
The coordinates are rather artificial than physical.
Thus we are led to an idea that
the framework of quantum mechanics on a manifold must include
a method to calculate the probability distribution directly.
We have introduced $ \nu $ in the definition (ii)
of section~\ref{sect:definition} to include such the method.
$ \nu( \psi, \psi )( x ) $ is interpreted as the probability density
with respect to the measure $ \di \mu ( x ) $
for finding the particle around $ x \in M $.
\par
We can also introduce another way to calculate the probability distribution.
Let $ ( M, \B, \mu ) $ be a Borel measure space
and let $ \Gm $ be a Hilbert space.
We call $ P $ a probability-measure operator on $ M $
when $ P $ satisfies the following conditions:
\begin{enum1}
\item	$ P $ is a spectral measure on $ ( M, \B, \mu ) $.
	Namely, for arbitrary $ D \in \B $,
	$ P(D) $ is a projection operator on $ \Gm $,
	and $ P $ satisfies
	\begin{eqnarray}
		&&
		P( \bigcup_{n=1}^\infty D_n ) = \sum_{n=1}^\infty P( D_n ),
		\quad D_n \in \B,
		\quad  D_m \cap D_n = \emptyset \, ( m \ne n ),
		\label{eqn:4.13}
		\\
		&&
		P( D_1 \cap D_2 ) = P( D_1 ) P( D_2 ) = P( D_2 ) P( D_1 ),
		\qquad D_1, D_2 \in \B,
		\label{eqn:4.14}
		\\
		&&
		P( M ) = I \quad ( \mbox{identity operator on } \Gm ).
		\label{eqn:4.15}
	\end{eqnarray}
\item	$ P $ and $ \mu $ are mutually absolutely continuous,
	namely, for $ D \in \B $
	\begin{equation}
		\mu( D ) = 0 \: \Longleftrightarrow \: P( D ) = 0.
		\label{eqn:4.16}
	\end{equation}
\end{enum1}
By virtue of (\ref{eqn:4.16}), for arbitrary $ \varphi, \psi \in \Gs $,
\begin{equation}
	\mu( D ) = 0
	\: \Longrightarrow \:
	\bra \varphi, P( D ) \psi \ket = 0
	\label{eqn:4.17}
\end{equation}
Therefore, according to the Radon-Nikod{\'y}m theorem,
there exists a unique $ \nu( \varphi, \psi ) \in \F( M ) $ such that
\begin{equation}
	\bra \varphi, P( D ) \psi \ket =
	\int_D \nu( \varphi, \psi )( x ) \, \di \mu ( x ),
	\qquad D \in \B
	\label{eqn:4.18}
\end{equation}
Moreover, (\ref{eqn:4.16}) also implies
\begin{equation}
	\mu( D ) \ne 0 \: \Longrightarrow \:
	\Gm_D := P( D ) \Gm \ne 0,
	\label{eqn:4.19}
\end{equation}
thus there exists $ \chi_D \in \Gm_D $ such that $ \chi_D \ne 0 $.
Since $ P( D^c ) \chi_D = 0 $, we have
\begin{equation}
	0 =
	\bra \chi_D, P( D^c ) \chi_D \ket =
	\int_{D^c} \nu( \chi_D, \chi_D )( x ) \, \di \mu ( x ),
	\label{eqn:4.20}
\end{equation}
which implies
$ \nu( \chi_D, \chi_D ) (x) = 0 \, ( x \notin D ) $.
Therefore we have shown that $ P $ reproduces $ \nu $ of the definition (ii)
in section~\ref{sect:definition}.
It is obvious that $ \nu $ also reproduces $ P $.
\subsubsection{Role of momentum operator}
As mentioned above, in quantum mechanics on $ \bm{R}^n $,
the momentum operators plays
the role of generators of the Lie algebra of translation.
Since they are self-adjoint, their exponentiations are unitary.
\par
We have included
the transformation group $ G $ rather than the Lie algebra $ \g $
directly in the definition in the section~\ref{sect:definition}.
The reason is that the action of $ G $ on $ M $ is directly described
in terms of points of $ M $ as $ \tau : G \times M \to M $,
while the action of $ \g $ on $ M $ is described
in terms of vector fields as $ \tau_* : \g \times M \to TM $,
and such a description complicates our framework.
\par
We have demanded that the action of $ G $ on $ M $ is transitive.
The reason is that we would like to make quantum mechanics
$ ( \Gm, \nu, \rho ) $ irreducible.
If the action is not transitive,
$ M $ is decomposed into $ G$-orbits
$ \{ M_\lambda \}_{ \lambda \in \Lambda } $,
and we can construct quantum mechanics
$ ( \Gm_\lambda, \nu_\lambda, \rho_\lambda ) $
for each orbit $ M_\lambda $.
\par
Moreover, we have demanded that the Borel measure $ \mu $ is $ G$-invariant.
This requirement may be too stringent.
By virtue of this requirement, our framework admits only
considerably narrow class of manifolds.
In other words, our consideration is restricted to proper homogeneous spaces.
\par
We have included
the local unitarity condition (\ref{eqn:2.10}) in the definition
to reflect $ G$-invariance of the measure $ \mu $.
This condition can be rephrased in terms of
the probability-measure operator $ P $ as follows:
\begin{equation}
	\rho(a) \, P(D) \, \rho(a)^\dagger = P(aD),
	\qquad a \in G, \quad D \in \B.
	\label{eqn:4.21}
\end{equation}
\subsection{Inhomogeneous spaces}
What we have investigated in this paper is essentially
representation of a Lie group $ G $
on a space of functions on a homogeneous space $ M $.
In general, we consider vector bundle-valued functions.
Examples we have studied are quantum mechanics on
the sphere, the torus and the projective space.
As other examples of homogeneous spaces,
the Stiefel manifold, the Grassmann manifold and the flag manifold are known,
and our scheme can be straightforwardly applied to them.
\par
On the other hand, we also know inhomogeneous spaces,
for example, the torus $ T_g $ of higher genus $ g \ge 2 $,
the Klein bottle and a manifold with boundary.
Apparently, our scheme cannot be applied to such manifolds.
For such the cases, we propose a naive substitution as follows.
If the manifold is a Riemannian manifold $ ( M, m ) $,
the space of complex-valued square-integrable functions on $ M $,
becomes a Hilbert space, which is denoted by $ \Gm = \Lb_{\:2}(M) $.
If $ M $ has boundary,
we impose a suitable boundary condition on the functions.
The Hamiltonian is defined by the Laplacian as $ H = - \frac12 \Delta_M $.
If it is needed, we may introduce a potential $ V(x) $ into $ H $,
as $ H = - \frac12 \Delta_M + V $.
However, this scheme does not have interesting structure such as
the commutation relations which the ordinary quantum mechanics possesses.
\subsection{Possibilities of extension}
Before closing this paper,
we would like to remark on some possibilities to extend our argument.
Non-trivial one of them is extension to quantum field theory.
For example,
the nonlinear sigma model treats a manifold-valued field~\cite{BKY}.
We shall explain the nonlinear sigma model briefly.
Assume that symmetry group $ G $ of a field theory
is spontaneously broken to its subgroup $ H $.
The vacua of broken phase form a homogeneous space $ M = G / H $,
which becomes a Riemannian manifold with $ G$-invariant metric $ m $.
Let $ ( N, n ) $ be the space-time with metric $ n $.
The lowest excitations of such a theory are massless modes
and are called Nambu-Goldstone bosons.
The nonlinear sigma model is an effective theory to describe
dynamics of Nambu-Goldstone bosons with field variable $ \phi : N \to M $,
that is a manifold-valued field.
The action of this model is defined by
\begin{eqnarray}
	S
	& = &
	\frac12 \int_N || \di \phi (x) ||^2 \, \di \nu (x)
	\nonumber \\
	& = &
	\frac12 \int_N \sum_{ \mu, \nu, a, b }
	n^{ \mu \nu }( x ) \,
	\partial_\mu \phi^a (x) \, \partial_\nu \phi^b (x) \,
	m_{ a b } ( \phi(x) ) \, \sqrt{ | n(x) | } \, \di^n x ,
	\label{eqn:4.22}
\end{eqnarray}
where $ \di \nu $ is the volume-measure defined by the metric $ n $.
This model is usually treated
by canonical quantization and by perturbative method.
However, we have already known that quantum mechanics on a manifold
has various inequivalent representations,
while quantum mechanics defined by canonical commutation relations
has a single representation.
Each inequivalent representation may give
different evaluation to a physical quantity.
Actually, we have known that
in the case of quantum mechanics on $ S^1 $,
there exist inequivalent representations parametrized by
$ \alpha \, ( 0 \le \alpha < 1 ) $,
and we have shown in another paper~\cite{Tani} that
both of energy spectrum and probability amplitude vary with $ \alpha $.
Observing quantum mechanics on a manifold,
we can expect that
the nonlinear sigma model also has various inequivalent representations
and exhibits different prediction for a physical quantity.
Unfortunately, by the perturbative method,
the global nature of $ M $ cannot be seen.
The global property is important to understand some aspects of field theories,
for example, the Wess-Zumino term and the soliton~\cite{Witten}.
Discussion on the global aspect of quantum theory is also given in~\cite{Tani}.
\par
As other possibilities of extension,
we are interested in path integral, statistics of identical particles and
relativistic quantum theory on a manifold.
%
%
\section*{Acknowledgments}
I am grateful to Prof. Y. Ohnuki and Prof. S. Kitakado
for suggestive discussions.
I would like to thank Prof. S. Kitakado for careful reading of my manuscript.
I am deeply indebted to Dr. Tsujimaru for his continuous encouragement.
%
%

%

\newpage
\begin{thebibliography}{99}
%
\bibitem{Bayen}
	F. Bayen, M. Flato, C. Fronsdal, A. Lichnerowicz and D. Sternheimer,
	{\it Ann. Phys.(N.Y.)} {\bf 111} (1978), 61; 111,
	and references cited therein.
\bibitem{Batalin}
	I.A. Batalin and I.V. Tyutin,
	{\it Nucl. Phys.} {\bf B345} (1990), 645
\bibitem{OK1}
	Y. Ohnuki and S. Kitakado,
	{\it Mod. Phys. Lett.} {\bf A7} (1992), 2477
\bibitem{OK2}
	Y. Ohnuki and S. Kitakado,
	{\it ``Fundamental Algebra for Quantum Mechanics on $ S^D $
	and Gauge Potentials''},
	Nagoya Univ. preprint DPNU-92-10 (1992);
	to be published in {\it J. Math. Phys.}
\bibitem{Tani}
	S. Tanimura,
	{\it ``Gauge Field, Parity and Uncertainty Relation
	of Quantum Mechanics on $ S^1 $''},
	Nagoya Univ. preprint DPNU-93-09 (1993);
	to be published in {\it Prog. Theor. Phys.}
\bibitem{Eguchi}
	T. Eguchi, P.B. Gilkey and A.J. Hanson,
	{\it Phys. Rep.} {\bf 66} (1980), 213
\bibitem{Nakahara}
	M. Nakahara,
	{\it ``Geometry, Topology and Physics''},
	Adam Hilger: Bristol and New York (1990)
\bibitem{Dictionary}
	Math. Soc. Japan ed.,
	{\it ``Encyclopedic Dictionary of Mathematics''},
	3rd ed. (in Japanese) Iwanami: Tokyo (1985);
	2nd ed. (translated to English) MIT Press (1987)
\bibitem{Ohnuki}
	Y. Ohnuki,
	{\it ``Unitary Representations of the Poincar{\'e} Group
	and Relativistic Wave Equations''},
	World Scientific: Singapore (1988), \S 5.4
\bibitem{Schulman}
	L. Schulman,
	{\it Phys. Rev.} {\bf 176} (1968), 1558
\bibitem{BKY}
	For a review,
	M. Bando, T. Kugo and K. Yamawaki,
	{\it Phys. Rep.} {\bf 164} (1988), 217
\bibitem{Witten}
	E. Witten,
	{\it Nucl. Phys.} {\bf B223} (1983), 422; 433
\end{thebibliography}
\end{document}